\documentclass[
 reprint,onecolumn,
 amsmath,amssymb,
 aps,
prb,
pdf]{revtex4-1}

\usepackage{epsfig,bm,epsf,graphics}
\usepackage{graphicx,subfigure}
\usepackage{dcolumn}
\usepackage{bm}

\begin{document}
\preprint{APS/123-QED}

\title{Skyrmion Crystal and Phase Transition in Magneto-Ferroelectric Superlattices: Dzyaloshinskii-Moriya Interaction in a Frustrated $J_1-J_2$ Model}

\author{Ildus F. Sharafullin $^{1}$ and H. T. Diep $^{2}$ }

\affiliation{
$^{1}$ \quad Bashkir State University, 32, Validy str, 450076, Ufa, Russia.\\
$^{2}$ \quad Laboratoire de Physique Th\'eorique et Mod\'elisation,
Universit\'e de Cergy-Pontoise, CNRS, UMR 8089, 2 Avenue Adolphe
Chauvin, 95302 Cergy-Pontoise, Cedex, France ; diep@u-cergy.fr}

\date{\today}

\begin{abstract}
The formation of a skyrmion crystal and its phase transition are studied taking into account the Dzyaloshinskii-Moriya (DM) interaction at the interface between a ferroelectric layer and a magnetic layer
in a superlattice. The frustration is introduced in both magnetic and ferroelectric
films.  The films have the simple cubic lattice structure. Spins inside  magnetic layers are Heisenberg spins interacting with each other via nearest-neighbor (NN) exchange $J^m$ and next-nearest-neighbor (NNN) exchange $J^{2m}$.  Polarizations in the ferroelectric layers are assumed to be of Ising type with NN and NNN interactions $J^f$ and $J^{2f}$. At the magnetoelectric interface, a DM interaction $J^{mf}$ between spins and polarizations is supposed.
The spin configuration in the ground state is calculated by the steepest descent method. In an applied magnetic field $\mathbf H$ perpendicular to the layers, we show that the formation of skyrmions at the magnetoelectric interface is strongly enhanced by the frustration brought about by the NNN antiferromagnetic interactions $J^{2m}$ and $J^{2f}$. Various physical quantities at finite temperatures are obtained by Monte Carlo simulations. We show the critical temperature, the order parameters of magnetic  and ferroelectric layers as functions of the interface DM coupling, the applied magnetic field  and $J^{2m}$ and $J^{2f}$.
The phase transition to the disordered phase is studied in details.

\end{abstract}

\maketitle




\section{Introduction}\label{sect-intro}

The localized topological spin-textures called "magnetic skyrmions"  currently attract intensive researches due to their fundamental properties and practical applications\cite{bogdanov1989thermodynamically,yu2010real,yu2011near, heinze2011spontaneous,romming2013writing,rosch2013skyrmions}. Indeed, skyrmions are promising structures for future digital technologies\cite{leonov2016chiral,moreau2016additive,soumyanarayanan2017tunable,dupe2016engineering,muller2016edge,rosch2017spintronics,shen2019spin,fert2013skyrmions,bessarab2019stability}. In addition, next-generation spintronic devices can be based on metastable isolated skyrmions guided along magnetic stripes\cite{fert2013skyrmions,tomasello2014strategy,koshibae2015memory,kang2016voltage}. Skyrmions have been experimentally observed in many materials, in particular in magnetic materials \cite{zhang2017skyrmion,muhlbauer2009skyrmion,yu2010real,heinze2011spontaneous,romming2013writing,du2015edge,jiang2015blowing, leonov2016chiral,leonov2016properties,woo2016observation,jiang2017direct,litzius2017skyrmion,woo2017spin}, multiferroic materials \cite{seki2012observation}, ferroelectric materials \cite{nahas2015discovery}, and semiconductors \cite{kezsmarki2015neel}.  Skyrmions have been seen also in helimagnets \cite{muhlbauer2009skyrmion,yu2010real}. Under an applied magnetic field, it has been shown that the helical spin configuration is transformed into a skyrmion lattice with a triangular structure \cite{DiepSD}. In a strong magnetic field the spin ordering is ferromagnetic (FM) but there exist isolated stable skyrmions as  topological defects \cite{butenko2010stabilization,rossler2011chiral}.

Real magnetic materials have complicated interactions and there are large families of frustrated systems such as the heavy lanthanides metals (holmium, terbium and dysprosium) \cite{zverev2014magnetic, zverev2015magnetic}, helical MnSi \cite{stishov2007magnetic}. Other interesting properties of skyrmions in  magnetically frustrated systems have also been  investigated\cite{leonov2017edge, lin2016ginzburg, hayami2016bubble, hayami2016vortices, lin2016magnetic, batista2016frustration, yuan2017skyrmions, rozsa2016skyrmions, rozsa2017formation, sutcliffe2017skyrmion,DiepSD}. Multiferroics and superlattices of multiferroics (for example $PZT/LSMO$ and $BTO/LSMO$) currently attract many research activities on these materials because of the coexistence of ferroelectric and magnetic orderings. A large number of investigations was devoted to the non-uniform states in magneto-ferroelectric superlattices both theoretically \cite{sharafullin2019dzyaloshinskii} and experimentally\cite{zheng2004multiferroic, bibes2008multiferroics,mathur2008materials,nan1994magnetoelectric,sergienko2006role,udalov2018coulomb,ortiz2014monte}. In Ref. \cite{janssen1994dynamics,janssen1982microscopic} Janssen et al. have proposed a new model for the interaction between polarizations and spins in magneto-ferroelectric superlattices. Using this model,  the dynamics and configuration of domain walls for the unidimensional case have been simulated. Li et al. \cite{li2002monte} have proposed an algorithm based on the Monte Carlo method for a two-dimensional (2D) lattice.
Recently, magneto-ferroelectric superlattices draw much of attention as magneto-electric (ME) materials. This is due to intrinsic magneto-electric effects stemming from the spin-orbit interaction \cite{sergienko2006role} as well as from the spin charge-orbital coupling \cite{pyatakov2018magnetoelectricity}. It has been shown that surface ME effects appear due to the charge and spin accumulation \cite{maruyama2009large,udalov2018coulomb}. The enhancement of magnetoelectric effect due to phase separation was shown in Ref. \cite{alberca2015phase}. Many microscopic interaction mechanisms for different materials have been suggested. Among these, we can mention the lone skyrmion-pair mechanism \cite{karthik2012site}, the ferroelectricity in manganites\cite{garcia2014geometric}, the  multiferroicity induced by the spiral spin ordering\cite{xiang2011general}, the off-center shifts in geometrically frustrated magnets\cite{balents2010spin}. In Ref. \cite{pei2017frustration} it was shown that magnetic frustration results in a phase competition, which is the origin of the magnetoelectric response.
A possible experimental realization of an isolated skyrmion as well as a skyrmion lattice has been suggested \cite{gobel2017antiferromagnetic}. In Ref. \cite{bessarab2019stability} the 2D Heisenberg exchange model with  Dzyaloshinskii-Moriya (DM) interaction is used to study the lifetime  of antiferromagnetic skyrmions.  Spin waves and skyrmions in magnetoelectric superlattices with a DM interface coupling have been studied \cite{sharafullin2019dzyaloshinskii}.
Yadav et al. \cite{yadav2016observation} have produced complex topologies of electrical polarizations, namely nanometer-scale vortex-antivortex structure, using the competition between charge, orbital and lattice degrees of freedom in superlattices of alternate lead titanate and strontium titanate films. They showed that the vortex state is the low-energy state for various superlattice periods.
In Ref. \cite{zhang2017skyrmion}, the authors have explored skyrmions and antiskyrmions in a 2D
frustrated ferromagnetic system with competing exchange
interactions based on the $J_{1}-J_{2}$ classical Heisenberg model
on a simple square lattice \cite{lin2016ginzburg} with the dipole-dipole interaction,
neglected in previous works \cite{leonov2015multiply,leonov2017edge}.
Dipole-dipole interaction has been shown to create the frustration in systems of skyrmions.
The interface-induced skyrmions have been investigated.
The superstructures are obviously the best to create interactions of
skyrmions on different interfaces causing very particular dynamics
compared to interactions between skyrmions of the same interface. We can  mention
a theoretical study of two
skyrmions on two-layer systems \cite{koshibae2017theory} using micromagnetic
modeling, and an analysis based on the Thiele equation. It has been found that there is  a reaction between them such as the collision and a bound
state formation. The dynamics of these skyrmions depends on the sign of DM
interaction,  and the number of  two
skyrmions is well described by the Thiele equation. Another interesting aspect is a colossal spin-transfer-torque effect of bound skyrmion pair revealed in antiferromagnetically coupled bilayer systems.  Note that the study of the current-induced motion using the Thiele equation was  carried in a skyrmion lattice through two models of the pinning
potential\cite{martinez2016topological}.
Shi-Zeng Lin et al \cite{Shi-ZengLin} have studied the dynamics of skyrmions in chiral magnets in the presence of a spin polarized current. They have also shown that skyrmions can be created by increasing the current in the magnetic spiral state.
Numerical simulations of Landau-Lifshitz-Gilbert equation have been performed in Ref. \cite{J-Iwasaki} which reveals a remarkably robust and universal current-velocity relation of the skyrmion motion driven by the spin-transfer-torque. This is unaffected by either impurities or non adiabatic effect, in sharp contrast to the case of domain wall or spin helix.

Note that in Ref. \cite{sharafullin2019dzyaloshinskii} we have studied  effects of Dzyaloshinskii-Moriya (DM) magneto-ferroelectric coupling in a superlattice formed by  "unfrustrated"  magnetic and ferroelectric films.
In zero field, we have shown that the GS spin configuration is periodically non collinear. By the use of a Green's function technique, we have calculated the spin-wave  spectrum in a monolayer and in a bilayer sandwiched between ferroelectric films. We have shown in particular that the DM interaction strongly affects the long-wavelength spin-wave mode.
In a magnetic field $\mathbf H$ applied in the $z$ direction perpendicular to the layers, we have shown that skyrmions are arranged  to form a crystalline structure  at the interface.

In this paper we study a superlattice composed of alternate "frustrated" magnetic films and "frustrated" ferroelectric films. The frustration due to competing interactions has been extensively investigated during the last four decades. The reader is referred to Ref. \cite{DiepFSS} for reviews on theories, simulations and experiments on various frustrated systems.  Our present aim is to investigate the effect of the frustration in the presence of the DM interaction at the magnetoelectric interface. It turns  out that the frustration gives rise to an enhancement of skyrmions created by the DM interaction in a field $\mathbf H$ applied perpendicularly to the films.
Monte Carlo simulations are carried out to study the skyrmion phase transition in the  superlattice as functions of the frustration strength.

The paper is presented as follows. In section \ref{MGS}\ we describe our model and show how to determine the ground-state spin structure. The skyrmion crystal is shown with varying frustration parameters. Section \ref{MC} is devoted to the presentation of the Monte Carlo results for the phase transition in the system as a function of the frustration, in the presence of the interface DM coupling. These results show that the skyrmion crystalline structure is stable up to a transition temperature $T_c$.  Concluding remarks are given in section \ref{Concl}.

\section{Model and Skyrmion Crystal}\label{MGS}
\subsection{Model}
\begin{figure}[h]
\begin{center}
\includegraphics[scale=0.25]{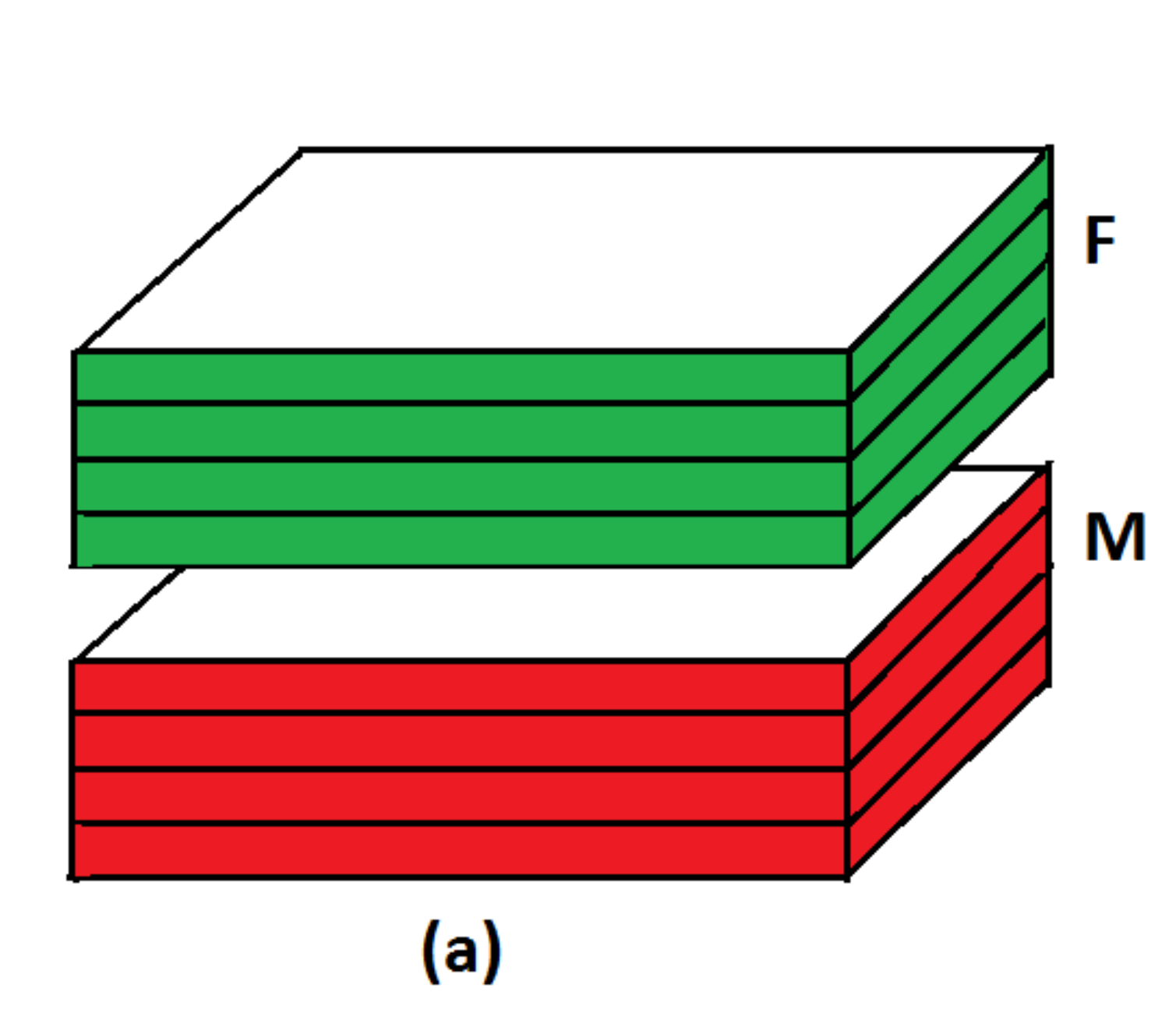}
\includegraphics[scale=0.35]{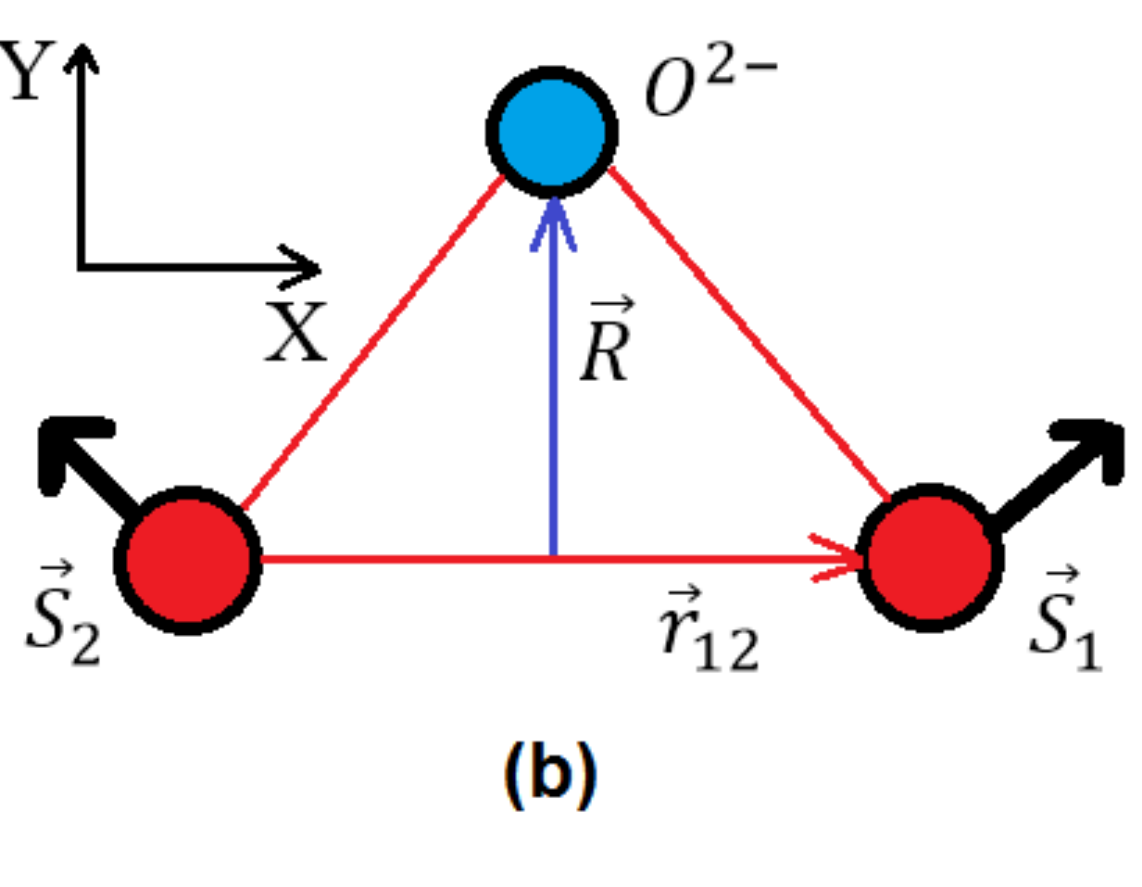}
\includegraphics[scale=0.35]{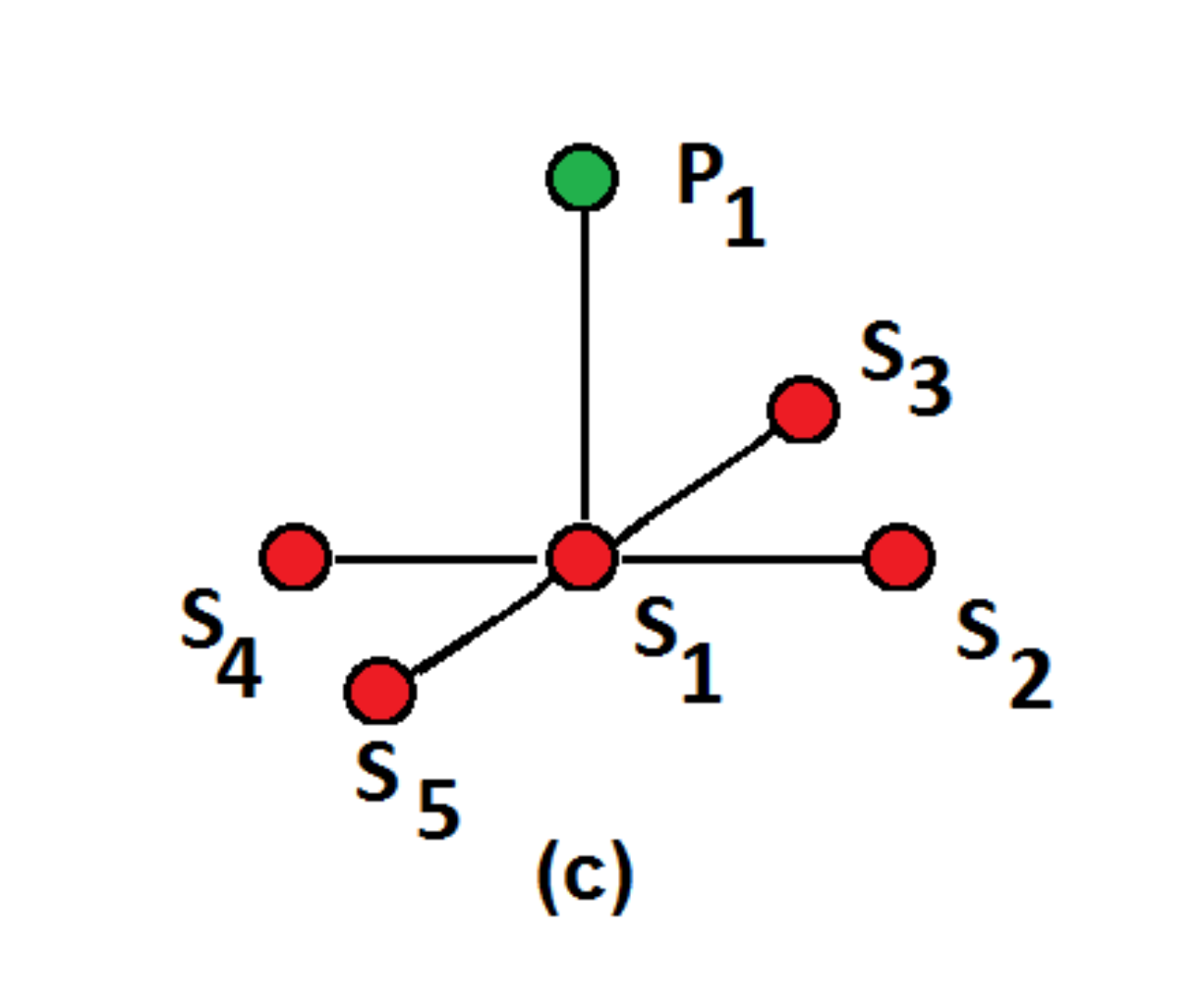}
\caption{(\textbf{a}) Magneto-ferroelectric superlattice, (\textbf{b}) Positions of the spins in the $xy$ plane and the position of non magnetic ion Oxygen, defining the DM vector (see text), (\textbf{c})  Interfacial coupling between a polarization $P$ with 5 spins in a Dzyaloshinskii-Moriya (DM) interaction, .}
\label{fig0}
\end{center}
\vspace{-10pt}
\end{figure}

The superlattice we study here is shown in Fig. \ref{fig0}a. It is composed of magnetic and ferroelectric films with   the simple cubic lattice. The Hamiltonian of
this magneto-ferroelectric superlattice is given by:

\begin{equation}\label{eq-ham-sysm-1}
{\cal H}=H_{m}+H_{f}+H_{mf}
\end{equation}
where $H_m$ and $H_f$ indicate the magnetic and
ferroelectric parts, respectively.   $H_{mf}$ denotes Hamiltonian
of magnetoelectric interaction at the interface of two adjacent
films. We are interested in the frustrated regime. Therefore we
consider  the following magnetic Hamiltonian with the Heisenberg
spin model

\begin{equation}\label{eq-ham-sysm-2}
H_{m}=-\sum_{i,j}{J^{m}_{ij} \mathbf {S}_{i}\cdot \mathbf
{S}_{j}}-\sum_{i,k}{J^{2m}_{ik} \mathbf {S}_{i}\cdot \mathbf
{S}_{k}}-\sum_{i}\mathbf {H}\cdot \mathbf{S}_{i}
\end{equation}
where $\mathbf S_{i}$  is the spin occupying the i-th lattice site, $\mathbf {H}$
denotes the  magnetic field applied along the $z$ axis and  $J_{ij}^{m}$ the magnetic
interaction between two spins $\mathbf S_i$ and $\mathbf S_j$. We shall take into account both the nearest neighbors (NN) interaction, denoted by $J^m$, and the next-nearest neighbor (NNN) interaction denoted by $J^{2m}$. We consider $J^{m}>0$
to be the same everywhere in the magnetic
film. To introduce the frustration we shall consider $J^{2m}<0$, namely antiferromagnetic
interaction, the same everywhere.   The interactions between spins and polarizations
at the interface are given below.

Before introducing the DM interface interaction, let us emphasize that the bulk $J_1-J_2$ model on the simple cubic lattice has been studied with Heisenberg spins \cite{DiepPinettes} and the Ising model \cite{DiepTai} where $J_1$ and $J_2$ are both antiferromagnetic ($<0$). The critical value $|J_2^c|=0.25 |J_1|$ above which the bipartite antiferromagnetic ordering changes into a frustrated ordering where a line is with spins up, and its neighboring lines are with spins down. In the case of $J^m>0$ (ferro), and $J^{2m}<0$, it is easy to show that the critical value where the ferromagnetic becomes antiferromagnetic is $J^{2m}_c=-0.5J^m$. Below this value, the antiferromagnetic ordering replaces the ferromagnetic ordering.

For the ferroelectric film, we assume that electric polarizations are of Ising model of magnitude 1,  pointing in
the $\pm z$ direction. The Hamiltonian reads
\begin{equation}\label{eq-ham-sysm-3}
H_{f}=-\sum_{i,j}{J_{ij}^{f} {\mathbf P}_{i}\cdot {\mathbf
P}_{j}}-\sum_{i,k}{J_{ik}^{2f} {\mathbf P}_{i}\cdot {\mathbf P}_{k}}
\end{equation}
where $\mathbf P_{i}$ is the polarization on the i-th lattice site,
$J_{ij}^{f}$ the interaction parameter between polarizations at sites $i$ and $j$. Similar to the magnetic subsystem we will
take the same $J_{ij}^{f}=J^f>0$ for all NN pairs, and
$J_{ij}=J^{2f}<0$ for NNN pairs.

We know that the DM interaction is written as
\begin{equation}\label{2}
H_{DM}=\mathbf {D}_{i,j}\cdot\mathbf {S_{i}}\times\mathbf {S_{j}}
\end{equation}
where $\mathbf {S_{i}}$ is the spin at the i-th magnetic site, while
$\mathbf{D}_{i,j}$ denotes the Dzyaloshinskii-Moriya vector which is defined
as  $\mathbf{R}\times\mathbf{r_{i,j}}$ where
$\mathbf{R}$ is the displacement of the
ligand ion (oxygen) and $\mathbf{r_{i,j}}$ the unit vector
along the axis connecting  $\mathbf S_i$ and $\mathbf S_j$ (see Fig.
\ref{fig0}b).
One then has
\begin{eqnarray}
\mathbf D_{i,j}&=&\mathbf R \times \mathbf r_{i,j}\nonumber\\
\mathbf D_{j,i}&=&\mathbf R \times \mathbf r_{j,i}=-\mathbf D_{i,j}
\end{eqnarray}
We define thus
\begin{equation}\label{eq4}
\mathbf D_{i,j}=D e_{i,j}\mathbf z
\end{equation}
where $D$ is a constant, $\mathbf z$ the unit vector on the $z$ axis, and $e_{i,j}=-e_{j,i}=1$.

For the magnetoelectric interaction at the interface, we choose the interface Hamiltonian following Ref. \cite{sharafullin2019dzyaloshinskii}:

\begin{equation}
H_{mf}= \sum_{i,j,k}J_{i,j}^{mf} e_{i,j}\mathbf {P_{k}}\cdot \left
[\mathbf {S_{i}}\times \mathbf {{S_{j}}} \right ]\label{Hmf0}
\end{equation}
where $\mathbf {P_{k}}$ is the polarization at the site $k$ of the ferroelectric interface layer,  while $\mathbf {S_{i}}$ and $\mathbf {S_{j}}$ belong to the interface magnetic layer (see Fig. \ref{fig0}c).
In this expression $J_{i,j}^{mf} e_{i,j}\mathbf {P_{k}}$ is defined as
the DM vector which is along the $z$ axis, given by Eq. (\ref{eq4}).
When summing the neighboring pairs $(i,j)$, attention should be paid on
the signs of $e_{i,j}$ and $\mathbf {S_{i}}\times \mathbf {{S_{j}}}$ (see example in Ref. \cite{sharafullin2019dzyaloshinskii}).

Hereafter, we suppose $J_{i,j}^{mf}=J^{mf}$ independent of $(i,j)$.

Since $\overrightarrow{P_k}$ is in the $z$
direction,  i. e. the DM vector is in the $z$ direction,  in the
absence of an applied field the spins  in the magnetic layers will lie in the $xy$ plane to minimize the interface interaction energy, according to Eq. (\ref{Hmf0}).

In Eq. (\ref{Hmf0}), the magnetoelectric interaction
$J^{mf}$ favors a non-collinear spin structure in competition with the
exchange interactions $J^m$ and $J^{2m}$ which favor collinear (ferro and antiferro) spin configurations.  In ferroelectric layers, only collinear, ferro- or antiferromagnetic ordering, is possible because of the assumed Ising model for the polarizations. Historical demonstration, the DM interaction was supposed small with respect to the exchange terms in the Hamiltonian. However, in superlattices the magnetoelectric interaction is necessary
to create non-collinear spin ordering.
It has been shown that Rashba spin-orbit coupling can lead to a strong DM interaction at the interface \cite{YangH2018,ManchonA2015}, where the broken inversion symmetry at the interface can change the magnetic states. The DM interaction has been identified as a key ingredient in the creation, stabilization of skyrmions and chiral domain walls.

\subsection{Ground state}
 From Eq. (\ref{Hmf0}) we see that the interface
interaction is minimum when $\mathbf {S_{i}}$ and $\mathbf {S_{j}}$
lie in the $xy$ interface plane and perpendicular to each other in the absence of exchange interactions. When the exchange interactions are turned on, the collinear configuration will
compete with the DM perpendicular configuration. This  results in a non-collinear configuration as will be shown below.

We note that when the magnetic film has more than one layer, the angle between NN spins in each magnetic
layer is different. The determination of the angles is analytically difficult. We have to recourse to the numerical method called "steepest
descent method" to minimize the energy to get the ground state (GS): we calculate the local field acting on a spin and align it in the direction of the local field. We go to another spin and do the same thing until all spins are visited. We repeat the operation many times until the total energy becomes minimum.

In the simulations, a sample size $N\times N\times L$ has been used, with the linear lateral size  $N=60$, and thickness $L= L_m +L_f$, where $L_{m}=L_{f}=4$ ($L_{m}$: magnetic layer's thickness, $L_{f}$: ferroelectric layer's thickness).  We use the periodic
boundary conditions in the $xy$ plane.

For simplicity, we take exchange parameters between
NN spins and NN polarizations  equal to 1, namely $J^{m}=J^{f}=1$, for the
simulations. We investigate the effects of the interaction parameters $(J^{2m},J^{2f})$ and
$J^{mf}$. We note that the steepest descent method calculates the
 GS down to the value
$J^{mf}=-1.25$. For values lower than this, the DM interaction is so strong that the spin-spin angle $\theta$ tends to $
\pi/2$ so that magnetic exchange terms are zero.



Now we consider a case with the frustrated regime with $(J^{2f},J^{2m})\in(-0.4,0)$, namely above the critical value -0.5 as mentioned above.

\begin{figure}[h]
\vspace{10pt}
\begin{center}
\includegraphics[scale=0.17]{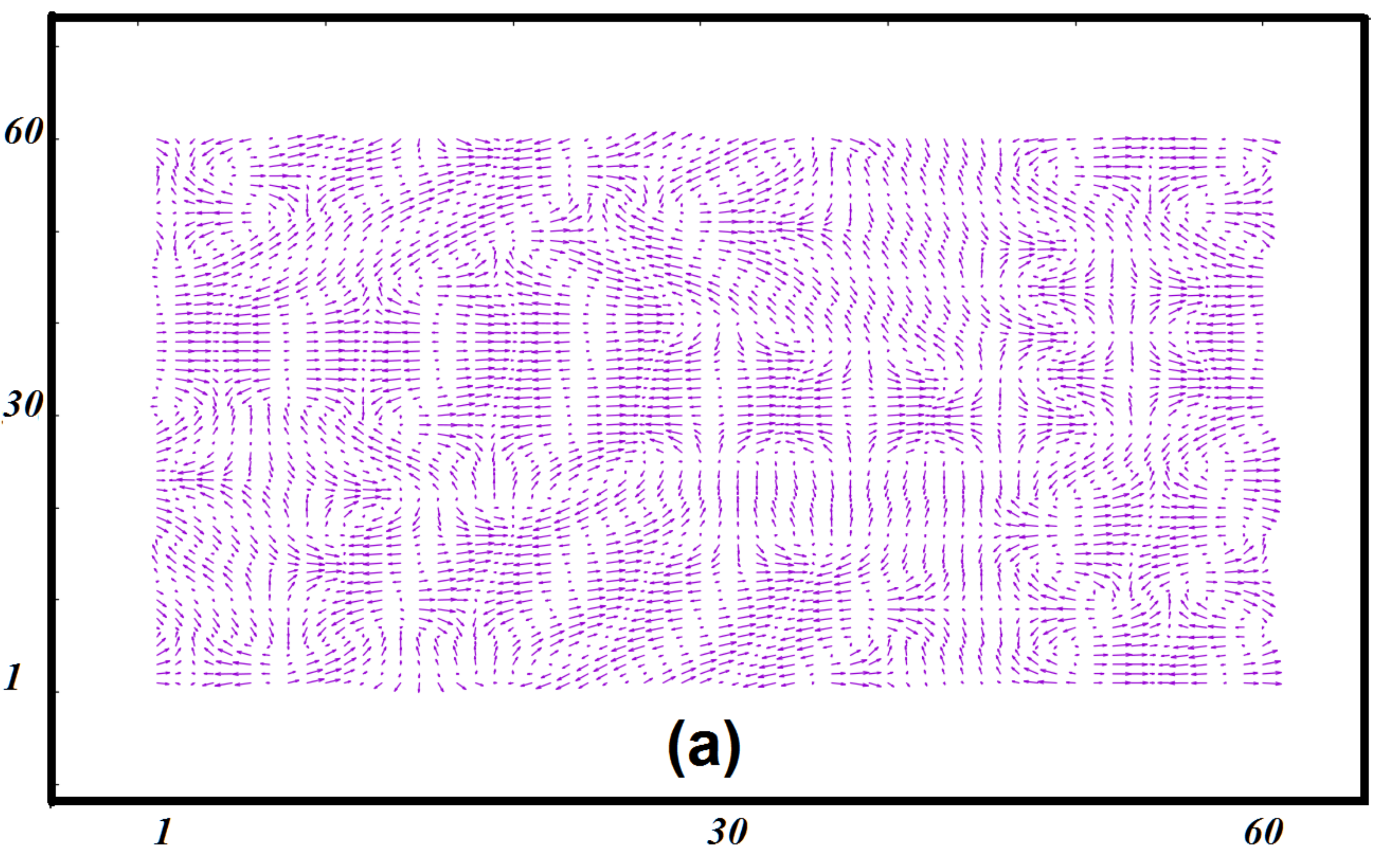}
\includegraphics[scale=0.40]{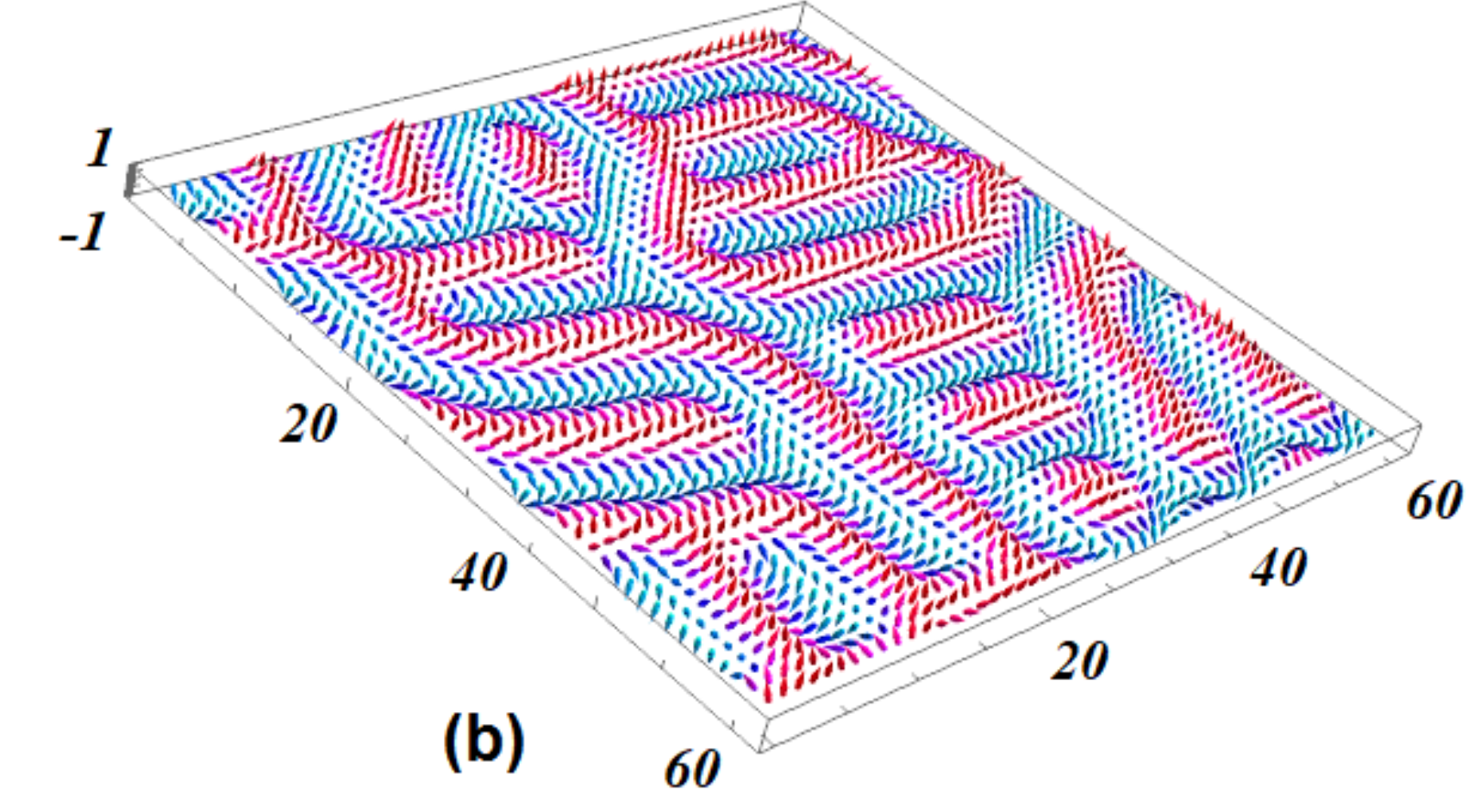}
\end{center}
\vspace{10pt} \caption{ (\textbf{a}) 2D view of the GS configuration of the
interface for $H=0$ with $J^{m}=J^{f}=1$, $J^{2m}=J^{2f}=-0.3$,
$J^{mf}=-1.25$, (\textbf{b}) 3D view. } \label{figH0} \vspace{10pt}
\end{figure}

The spin configuration in the case where $H=0$ is shown in Fig. \ref{figH0} for the interface magnetic layer. We observe here a stripe phase with long islands and domain walls. The inside magnetic layers have the same texture.

\begin{figure}[h]
\vspace{10pt}
\begin{center}
\includegraphics[scale=0.30]{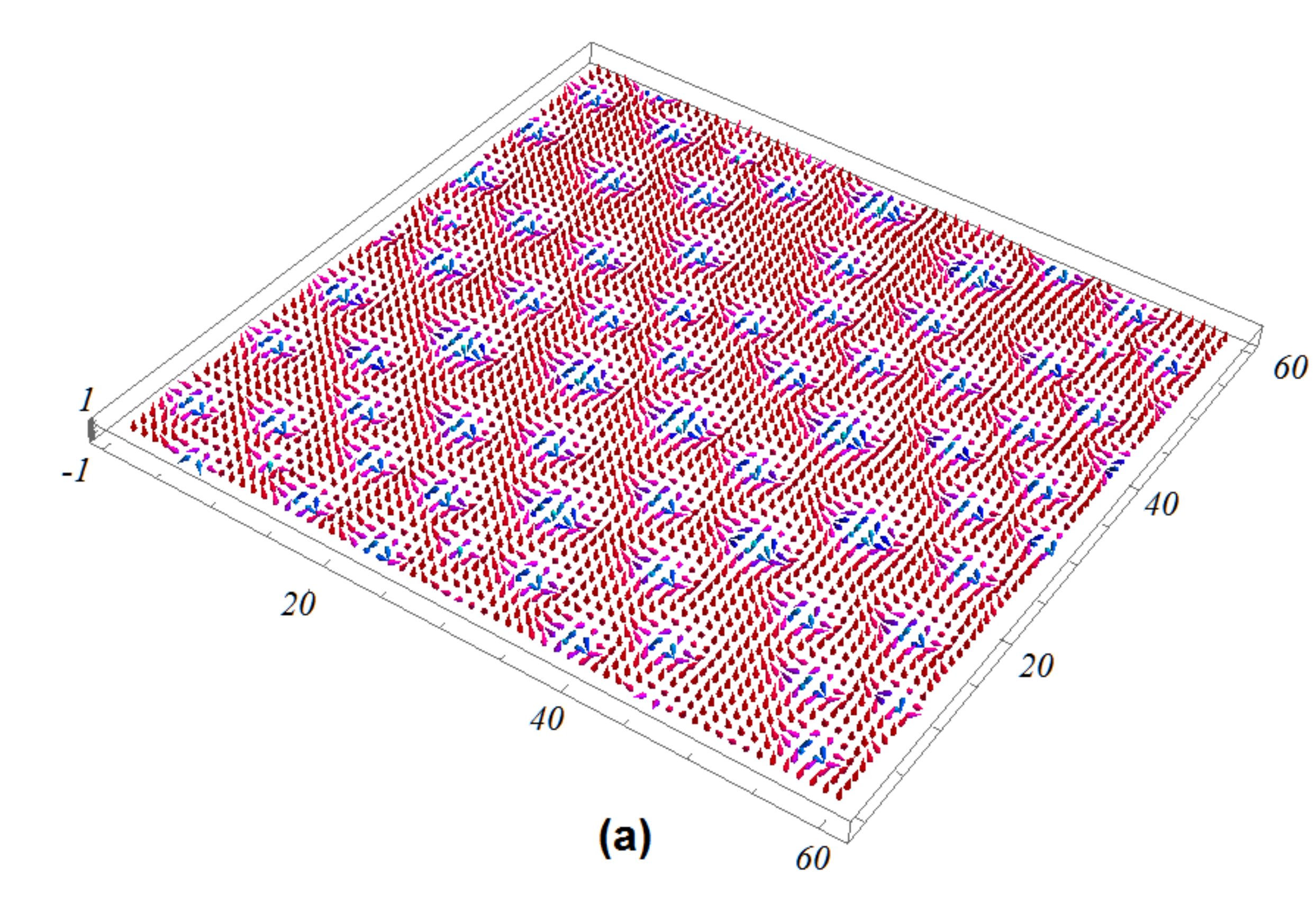}
\includegraphics[scale=0.30]{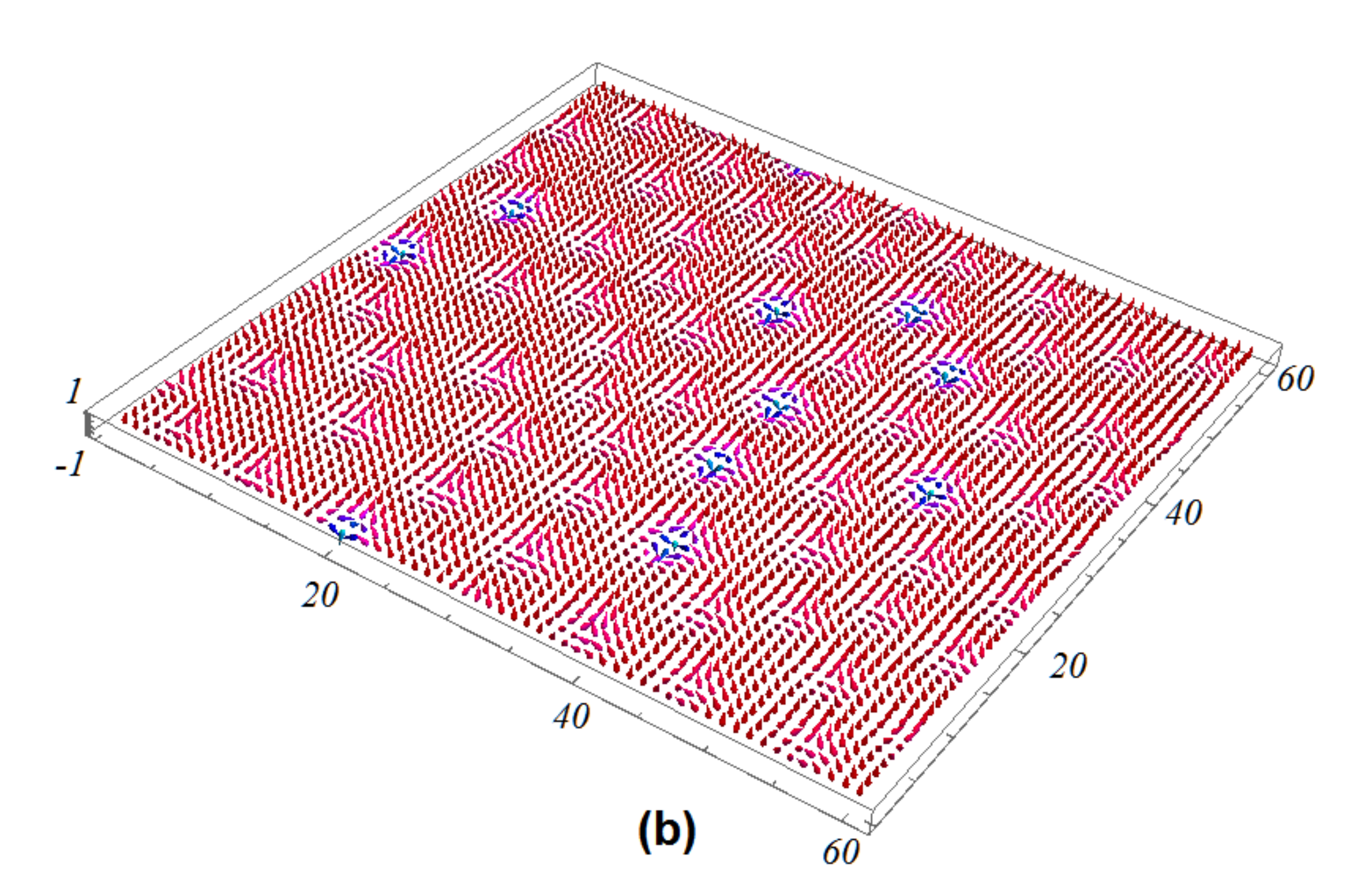}
\includegraphics[scale=0.35]{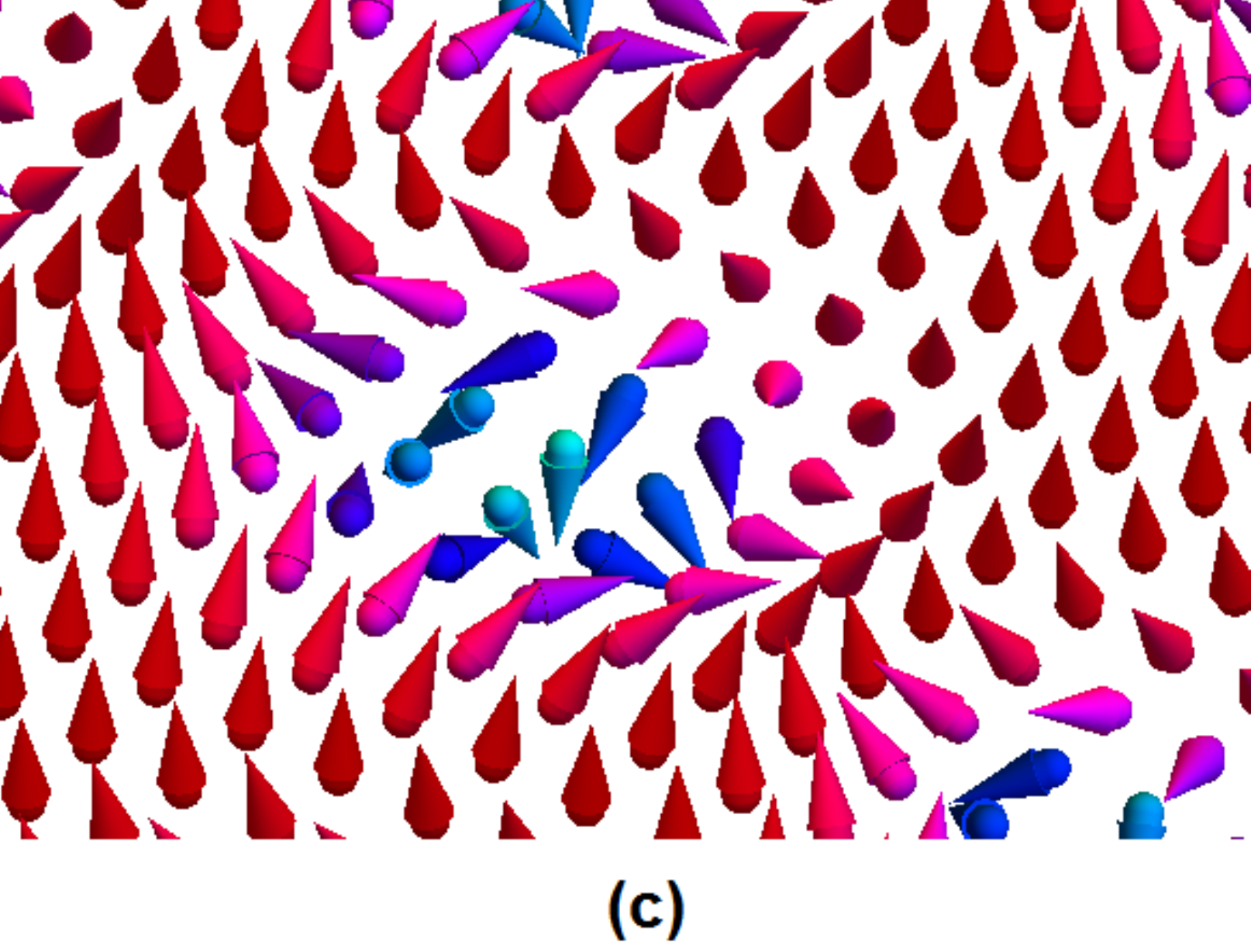}
\includegraphics[scale=0.18]{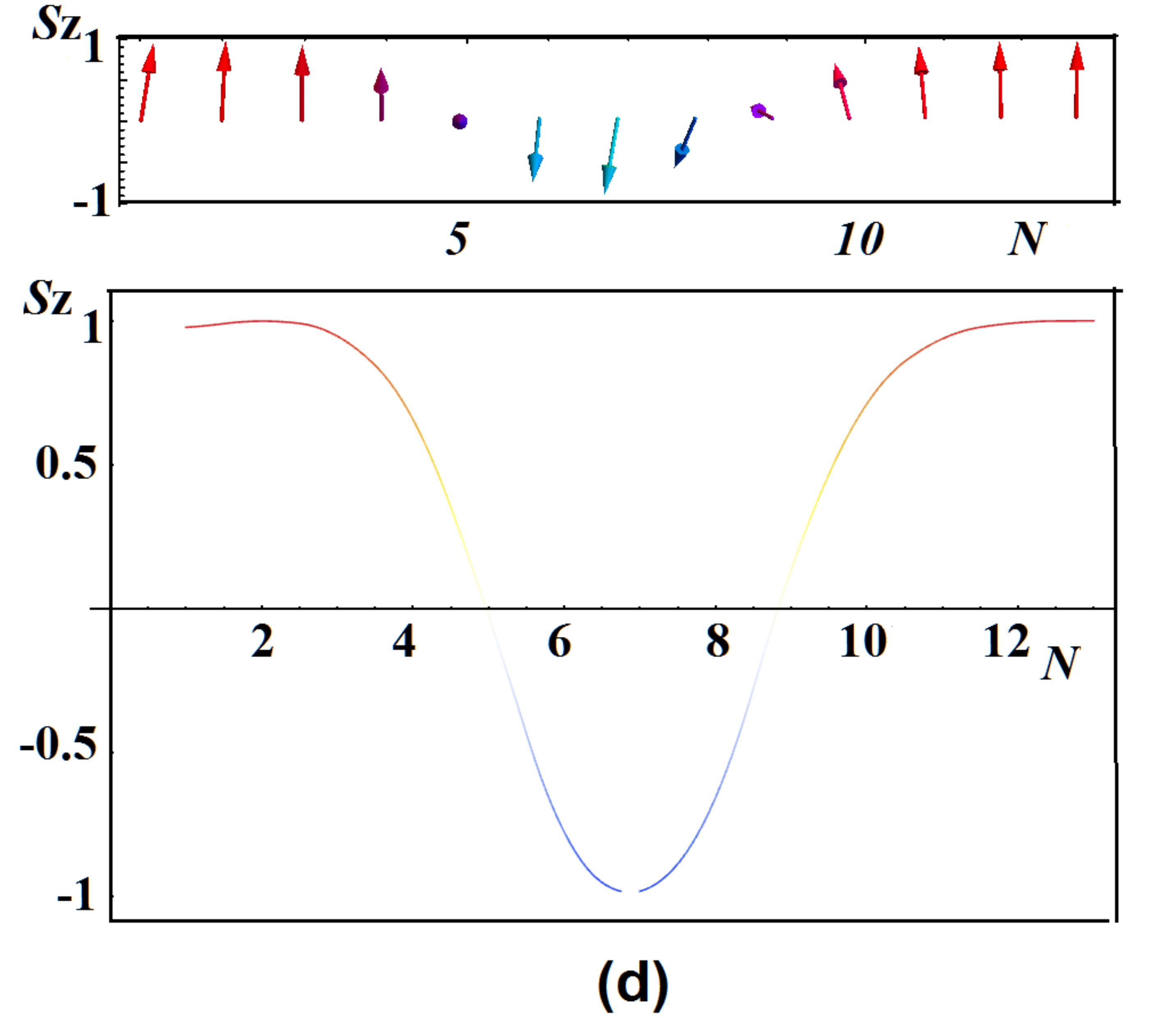}
\end{center}
\vspace{10pt} \caption{ (\textbf{a}) 3D view of the GS configuration of the
interface for moderate frustration $J^{2m}=J^{2f}=-0.2$.  , (\textbf{b}) 3D view of the GS structure of the interior magnetic layers, (\textbf{c}) zoom of a skyrmion on the interface layer: red denotes up spin, four spins with clear blue color are down spin, other colors correspond to spin orientations between the two. The skyrmion is of the Bloch type, (\textbf{d}) $z$-components of spins across the skyrmion shown in (c). Other parameters: $J^{m}=J^{f}=1$,
$J^{mf}=-1.25$ and $H=0.25$} \label{fig1} \vspace{10pt}
\end{figure}

When $H$ is increased, we observe the skyrmion crystal as seen in
Fig. \ref{fig1}: the GS
configuration of the interface  and beneath
magnetic layers obtained for $J^{mf}= -1.25$, with
$J^{2m}=J^{2f}=-0.2$ and external magnetic field $H=0.25$.  A zoom of a skyrmion shown in Fig. \ref{fig1}c and the $z$-components across a skyrmion shown in Fig. \ref{fig1}d indicate that the skyrmion is of Bloch type.

\begin{figure}[h]
\vspace{10pt}
\begin{center}
\includegraphics[scale=0.30]{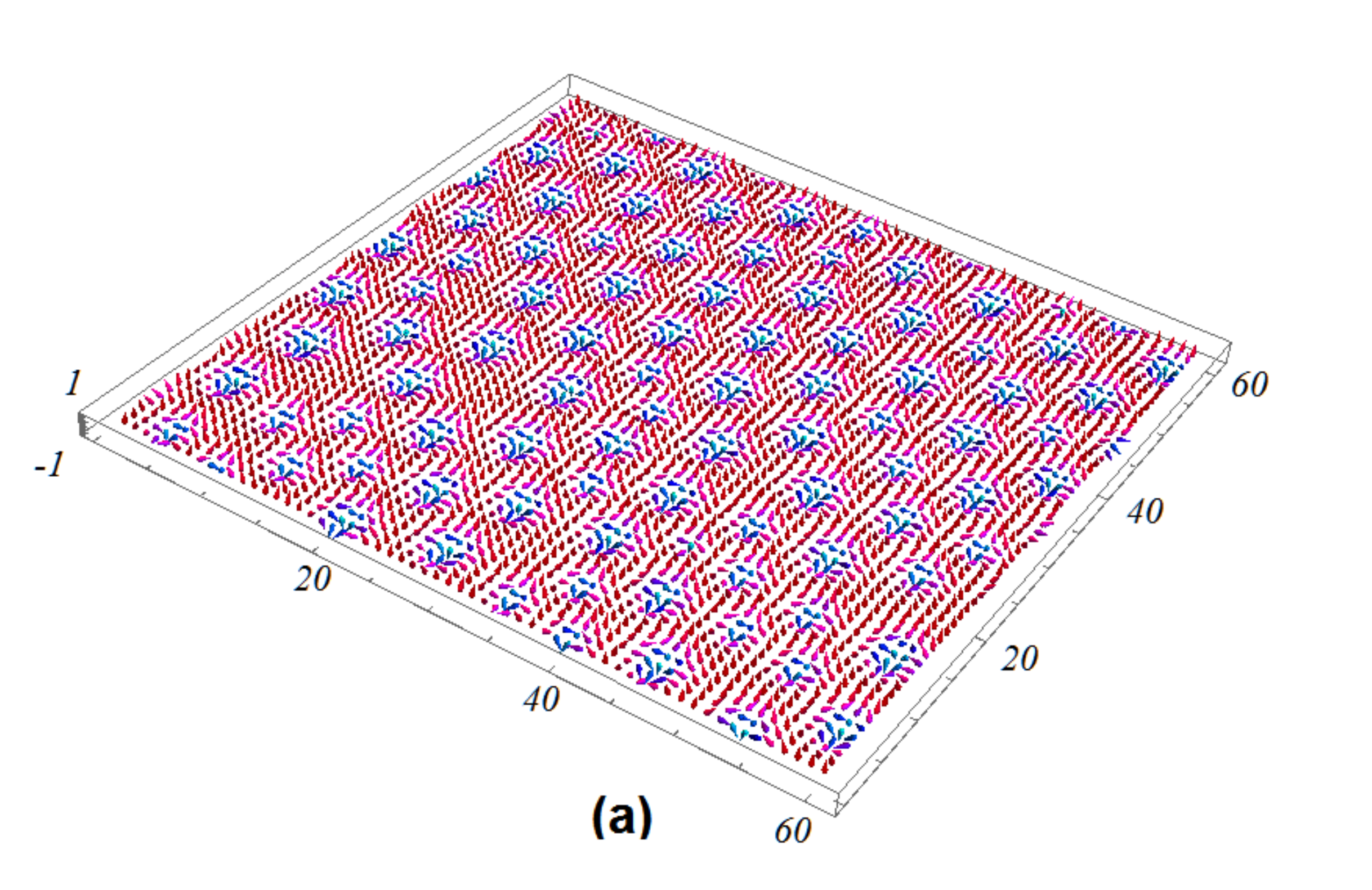}
\includegraphics[scale=0.30]{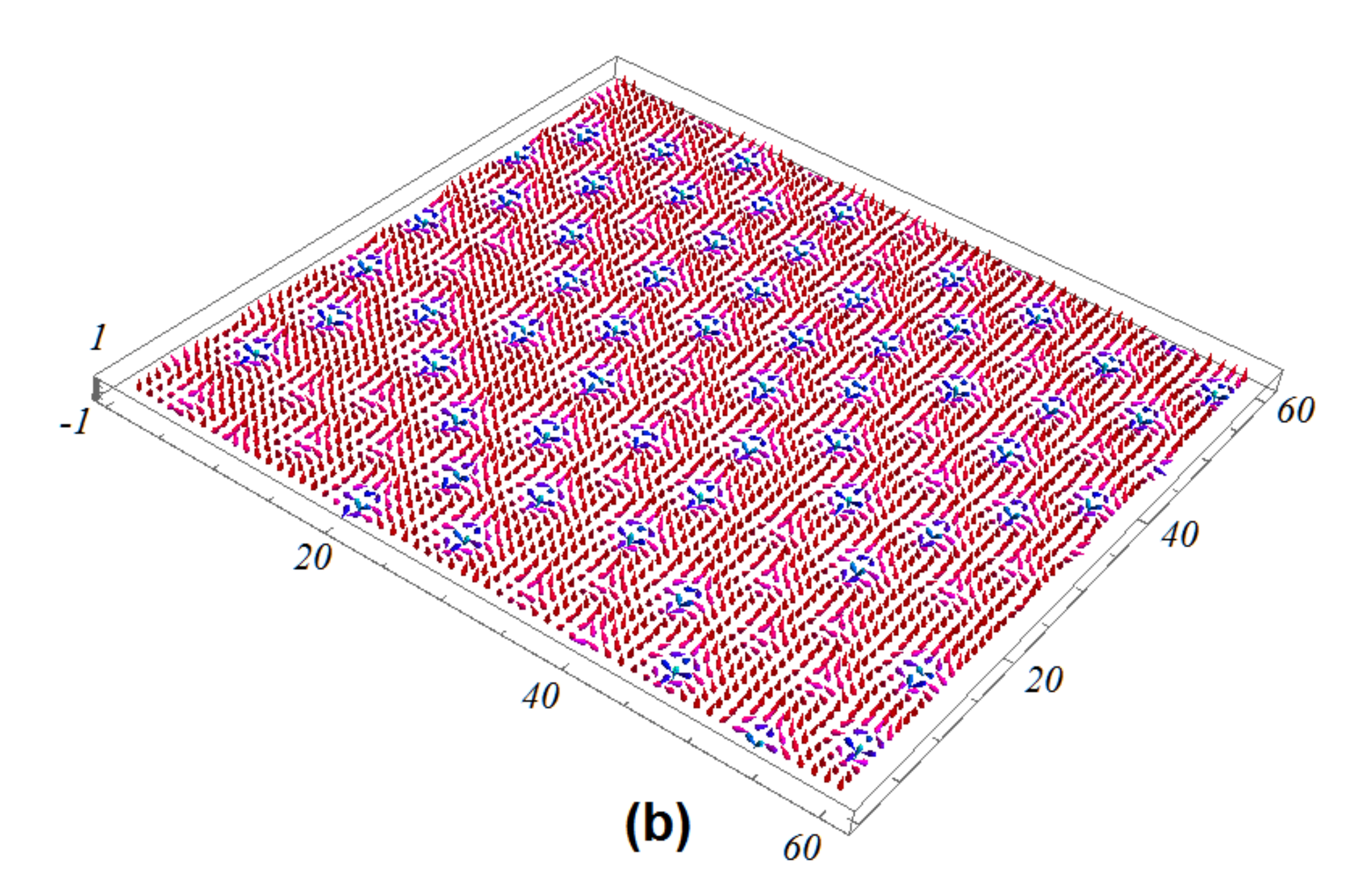}
\end{center}
\vspace{10pt} \caption{  3D view of the GS configuration of (\textbf{a}) the
interface, (\textbf{b}) the second layer,   for stronger frustration $J^{2m}=J^{2f}=-0.3$.    $J^{m}=J^{f}=1$, $J^{2m}=J^{2f}=-0.3$, $J^{mf}=-1.25$ and $H=0.25$. } \label{fig2} \vspace{10pt}
\end{figure}

At this field strength $H=0.25$, if we increase the frustration, for example $J^{2m}=J^{2f}=-0.3$, then the skyrmion structure is enhanced: we can observe a clear $3D$ skyrmion crystal structure not only in the interface
layer but also in the interior layers.  This is shown in Fig. \ref{fig2} where the interface  and the second layer are displayed.

\begin{figure}[h]
\vspace{10pt}
\begin{center}
\includegraphics[scale=0.30]{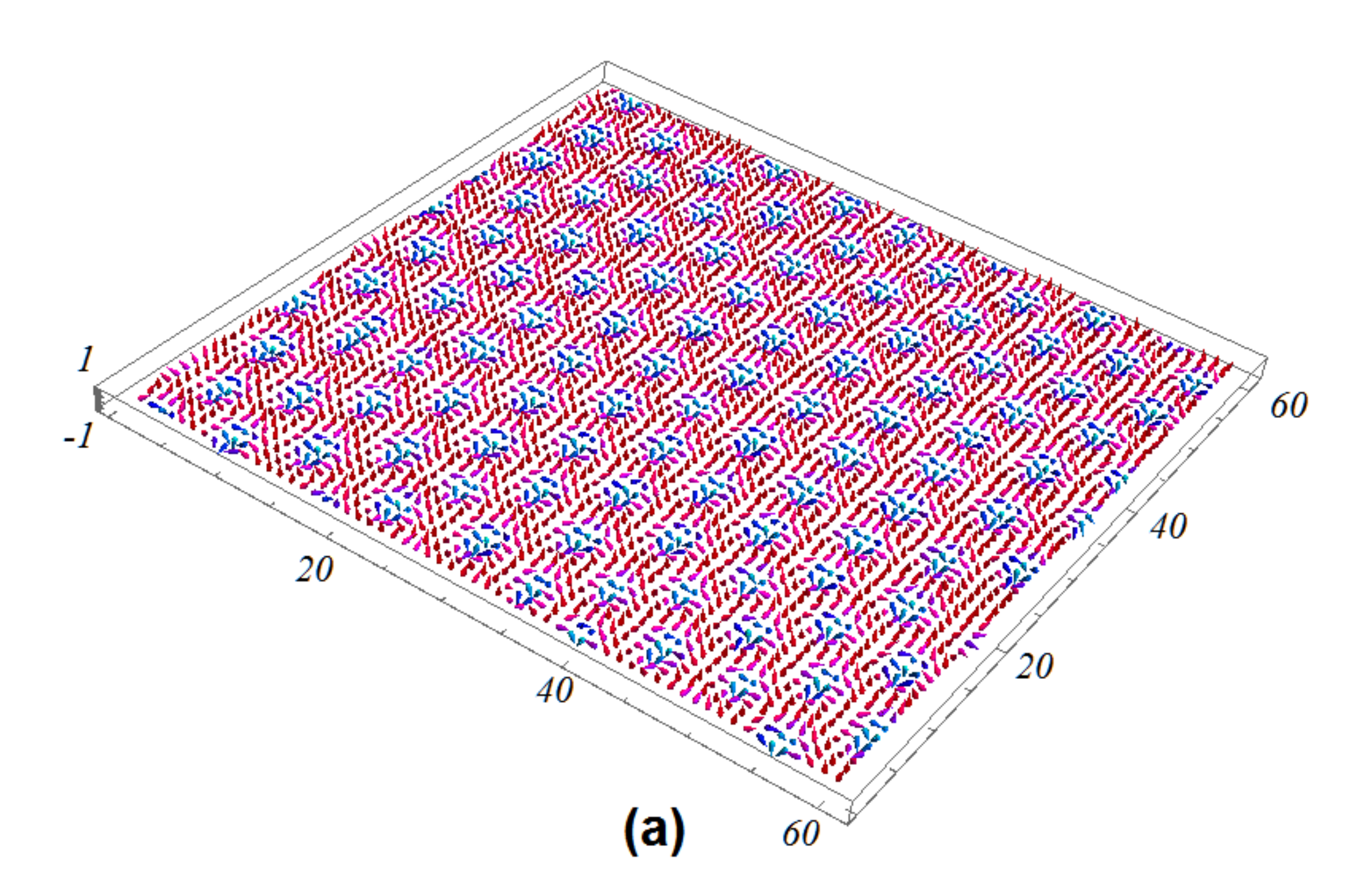}
\includegraphics[scale=0.30]{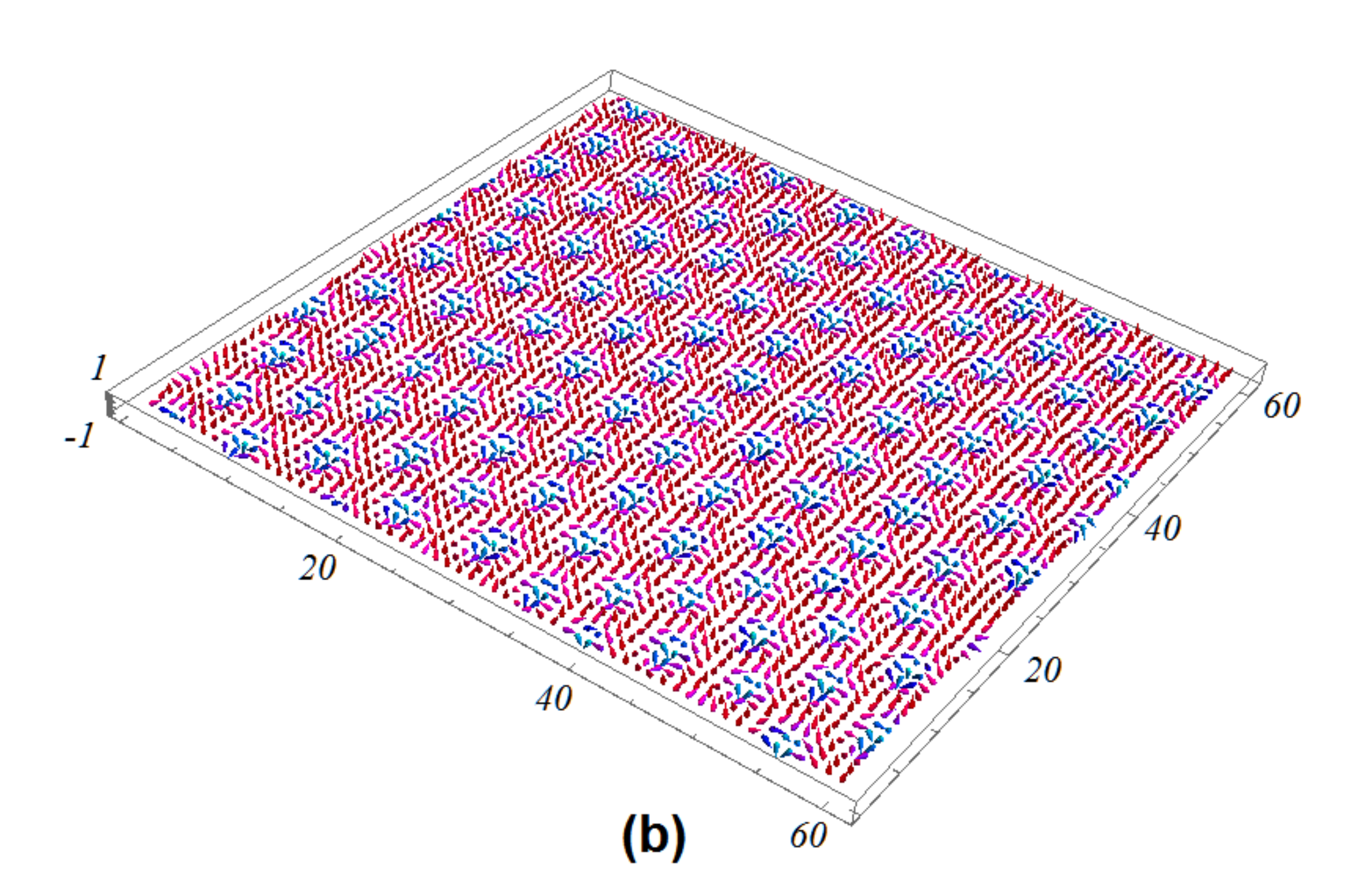}
\end{center}
\vspace{10pt} \caption{ Strongest frustration $J^{2m}=J^{2f}=-0.4$ (\textbf{a}) 3D view of the GS configuration of the
interface, (\textbf{b}) 3D view of the GS configuration of the second magnetic layers. Other parameters
$J^{m}=J^{f}=1$, $J^{mf}=-1.25$ and $H=0.25$.} \label{fig3} \vspace{10pt}
\end{figure}

The highest value of frustration where the skyrmion structure can be observed is when $J^{2m}=J^{2f}=-0.4$ close to the critical value -0.5. We show this case in Fig. \ref{fig3}: the GS configuration of the interface (a) and
second (interior) (b) magnetic layers are presented.  Other parameters are the same as in the previous figures, namely  $J^{mf}=
-1.25$ and $H=0.25$. We can observe a clear $3D$ skyrmion crystal
structure in the whole magnetic layers, not only near the interface layer. Unlike the
case where we do not take into account the interaction between $NNN$
\cite{sharafullin2019dzyaloshinskii}, in the present case where the frustration is very strong we see that a large number of skyrmions are
distributed over the whole magnetic layers with a certain periodicity close to a perfect crystal.

\begin{figure}[h]
\vspace{10pt}
\begin{center}
\includegraphics[scale=0.30]{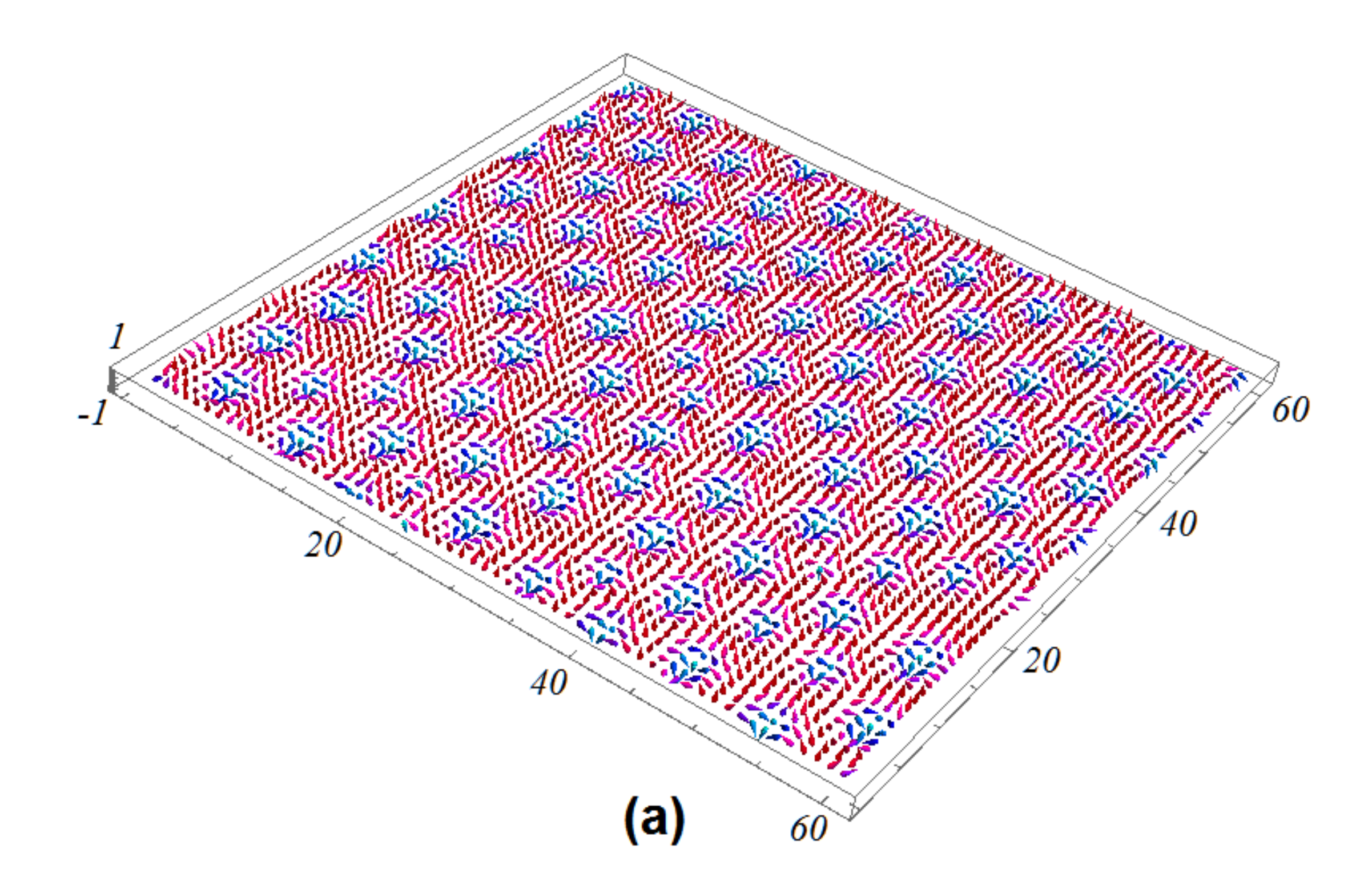}
\includegraphics[scale=0.30]{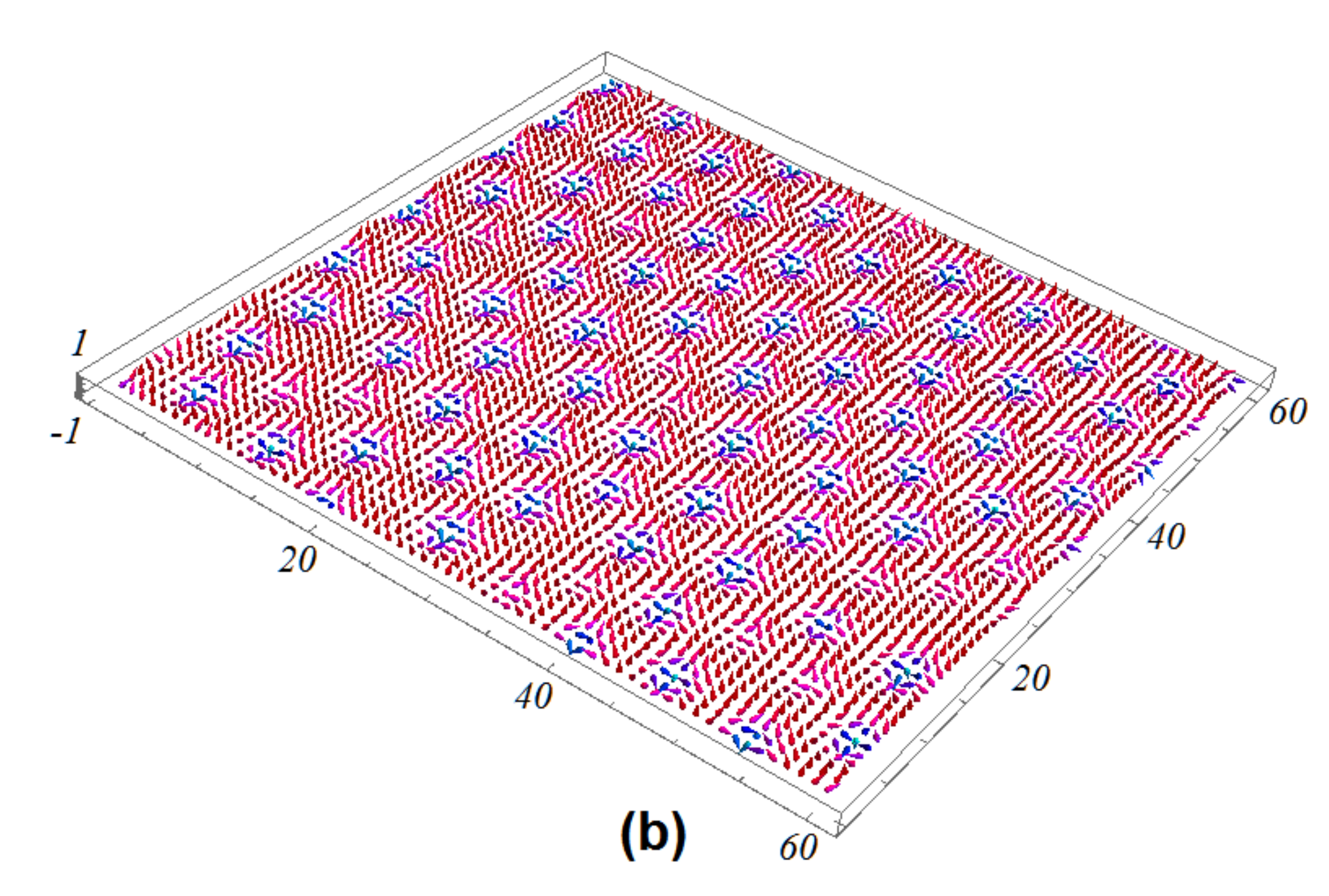}
\end{center}
\vspace{10pt} \caption{ (\textbf{a}) 3D view of the GS configuration of the
interface
 for $J^{m}=J^{f}=1$,$J^{2m}=-0.3, J^{2f}=-0.1$, $J^{mf}=-1.25$ and $H=0.25$, (\textbf{b}) 3D view of the GS configuration of the second magnetic layers,
 for  $J^{2m}=-0.3$, $J^{2f}=-0.1$. Other parameters: $J^{m}=J^{f}=1$, $J^{mf}=-1.25$ and $H=0.25$. } \label{fig4} \vspace{10pt}
\end{figure}

Though we take the same value for $J^{2m}$ and $J^{2f}$ in the figures shown above, it is obvious that only the magnetic frustration $J^{2m}$  is important for the skyrmion structure. The ferroelectric frustration affects only the stability of the polarizations at the interface.  As long as $J^{2f}$ does not exceed -0.5, the skyrmions are not affected by $J^{2f}$.
We show in Fig. \ref{fig4}  the GS configuration of the interface and the second
 magnetic layers for $J^{2m}=-0.3$ and
$J^{2f}=-0.1$ (other parameters:  $J^{mf}=-1.25$, $H=0.25$). We see that the skyrmion structure is not  different from the case ($J^{2m}=J^{2f}=-0.3$) shown in Fig. \ref{fig2}.

\begin{figure}[h]
\vspace{10pt}
\begin{center}
\includegraphics[scale=0.30]{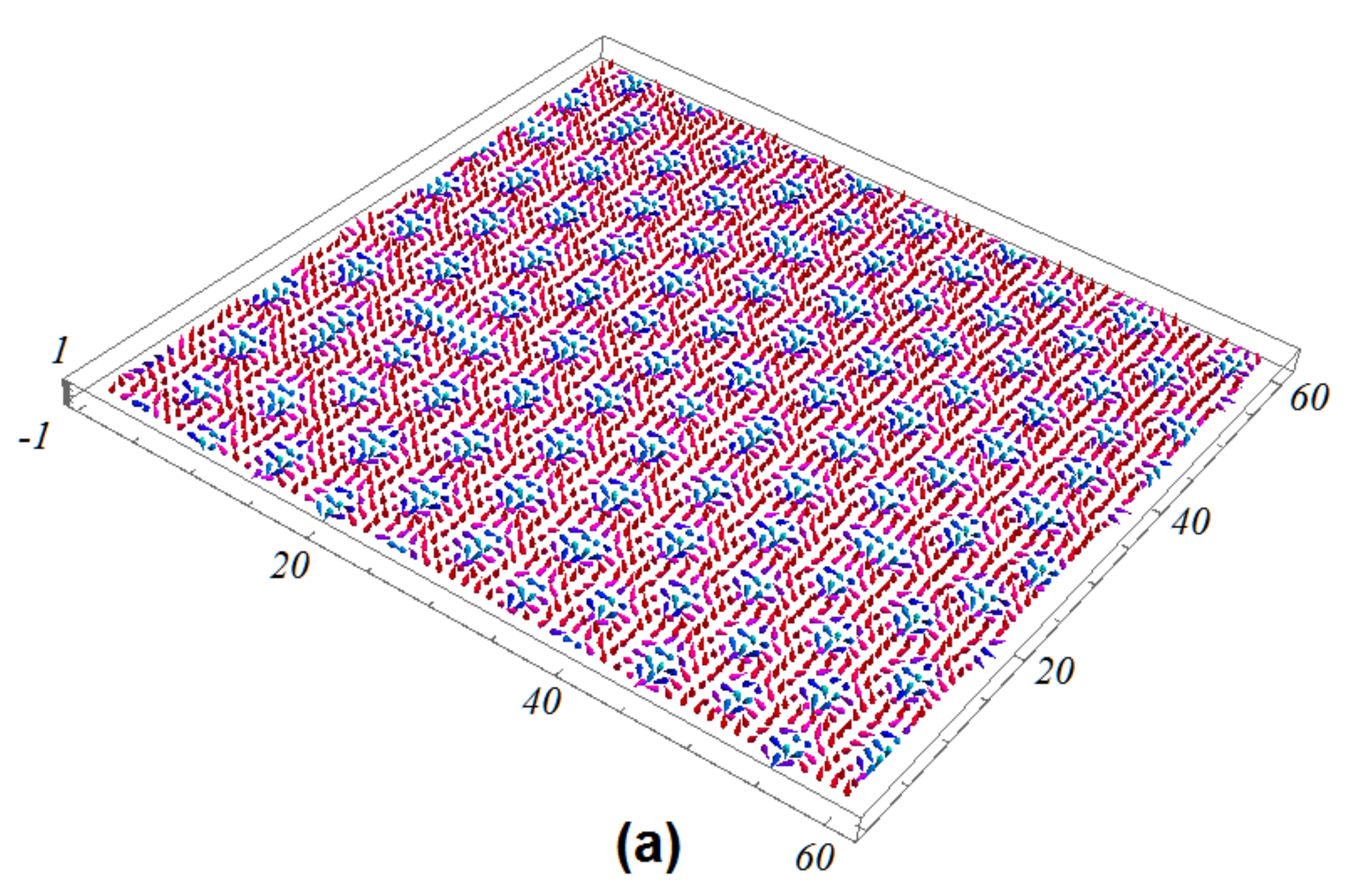}
\includegraphics[scale=0.30]{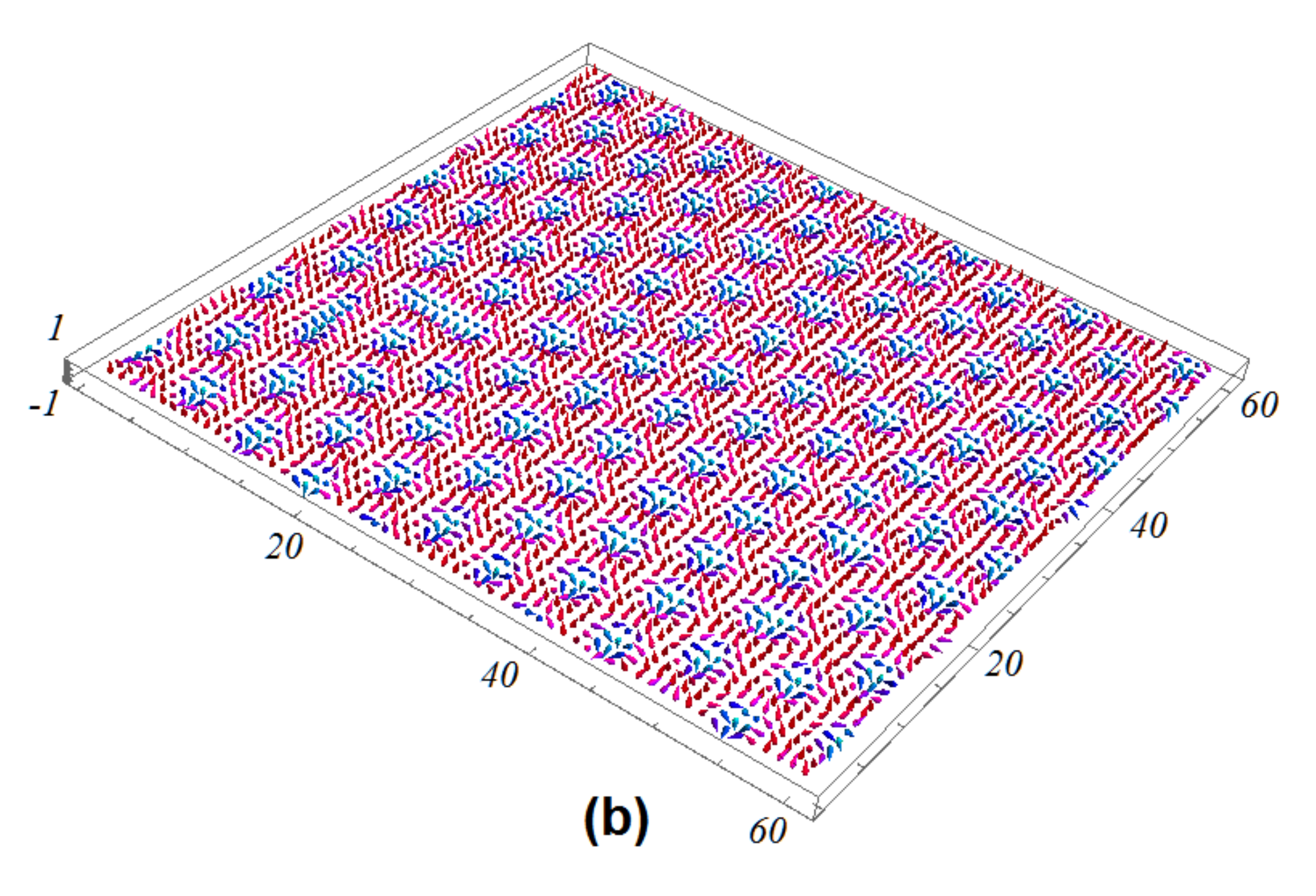}
\end{center}
\vspace{10pt} \caption{ (\textbf{a}) 3D view of the GS configuration of the
interface  $J^{2m}=-0.4$, $J^{2f}=-0.1$, (\textbf{b}) 3D view of the GS configuration of the interface for $J^{2m}=-0.4, J^{2f}=0$. Other parameters: $J^{m}=J^{f}=1$, $J^{mf}=-1.25$ and $H=0.25$. } \label{fig5} \vspace{10pt}
\end{figure}

Fig. \ref{fig5}  shows the GS configuration of the interface and second
(interior) magnetic layers for ($J^{2m}=-0.4, J^{2f}=-0.1$ and $J^{2f}=0$) which is not visibly different from  the case ($J^{2m}=J^{2f}=-0.4$) shown in Fig. \ref{fig3}.
We conclude here that when the magnetic frustration is strong enough, the ferroelectric frustration does not affect the skyrmion structure.

%

\begin{figure}[h]
\vspace{10pt}
\begin{center}
\includegraphics[scale=0.30]{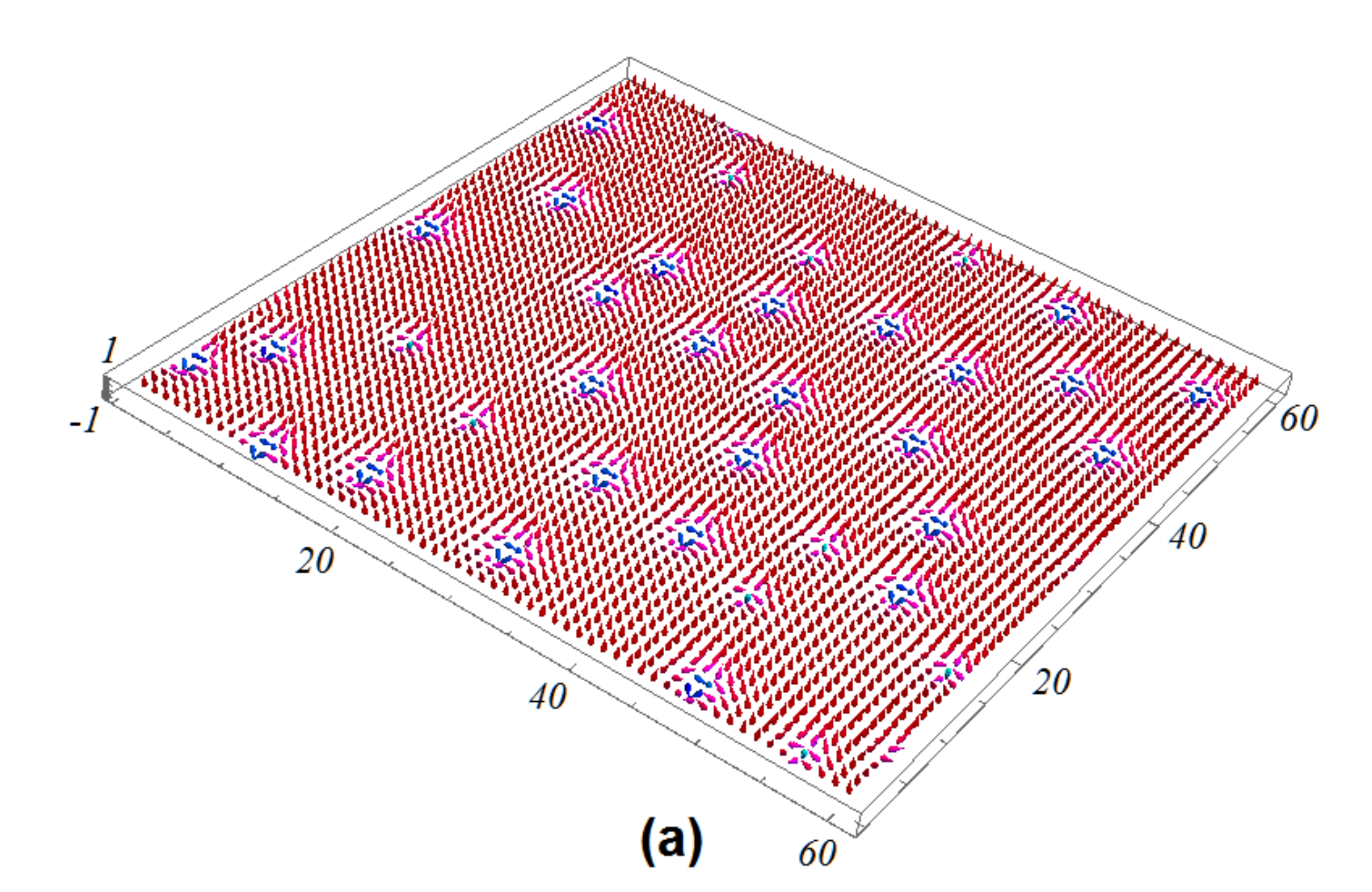}
\includegraphics[scale=0.27]{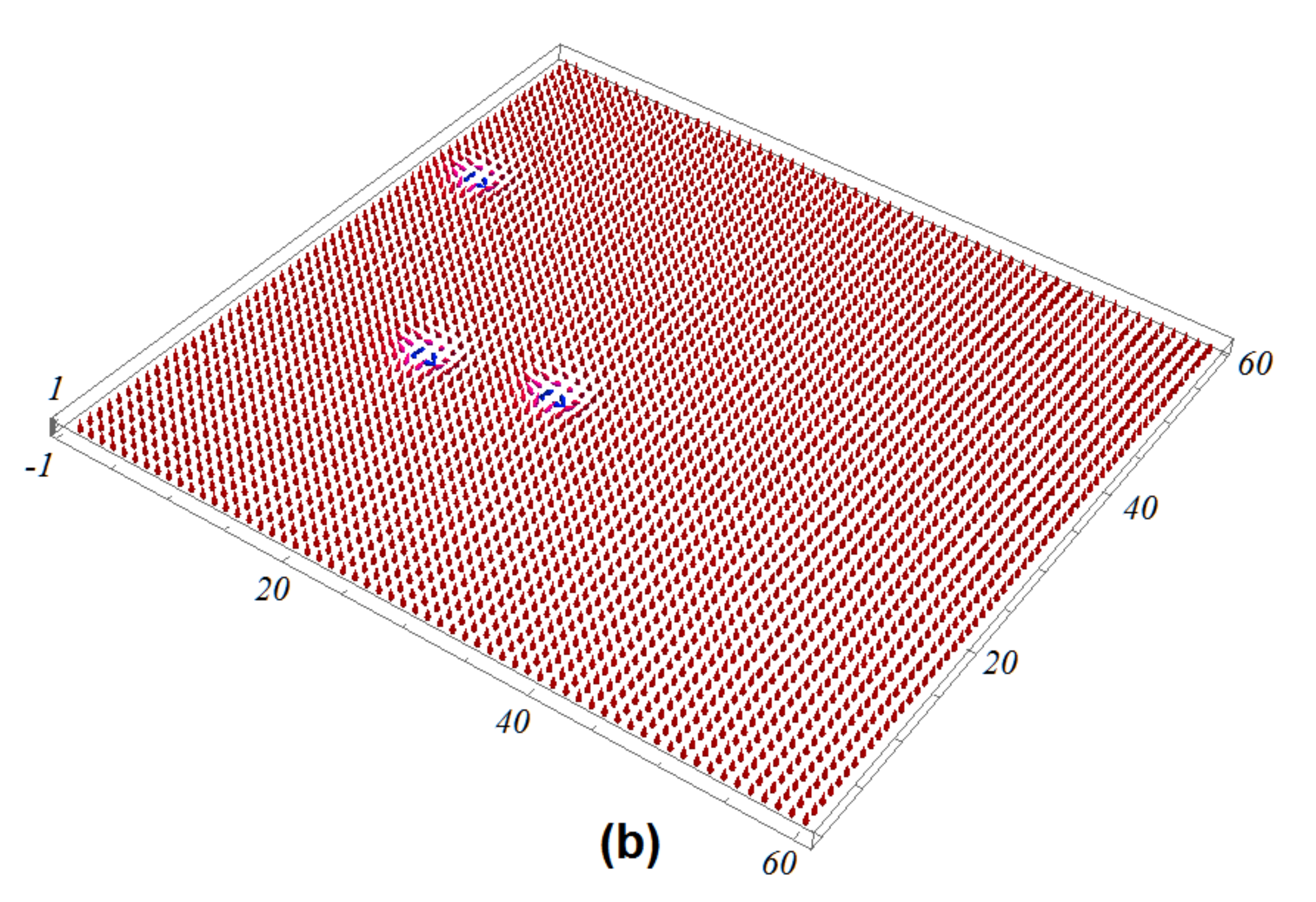}
\end{center}
\vspace{10pt} \caption{ (\textbf{a}) 3D view of the GS configuration of the
interface for $J^{2m}=-0.1$,
$J^{2f}=-0.3$, (\textbf{b}) 3D view of the GS configuration of the interface
for $J^{2m}=-0.1$, $J^{2f}=-0.4$. Other parameters: $J^{m}=J^{f}=1$,  $J^{mf}=-1.25$, $H=0.25$.} \label{fig7} \vspace{10pt}
\end{figure}

Now, if the magnetic frustration is not strong enough, the ferroelectric frustration plays an important role: Fig. \ref{fig7}a shows the GS configuration of the interface magnetic
layer for  ($J^{2m}=-0.1$, $J^{2f}=-0.3$) and Fig. \ref{fig7}a shows the case
of ($J^{2m}=-0.1$, $J^{2f}=-0.4$).   We see that skyrmions disappear when $J^{2f}=-0.4$.  Comparing this to the case ($J^{2m}=-0.4$, $J^{2f}=-0.4$) where skyrmions are clearly formed, we conclude that while magnetic frustration $J^{2m}$ enhances the formation of skyrmions, the ferroelectric frustration $J^{2f}$ in the weak magnetic frustration tends to suppress skyrmions.  The mechanism of these parameters when acting together seems to be very complicated.

\begin{figure}[h]
\vspace{10pt}
\begin{center}
\includegraphics[scale=0.30]{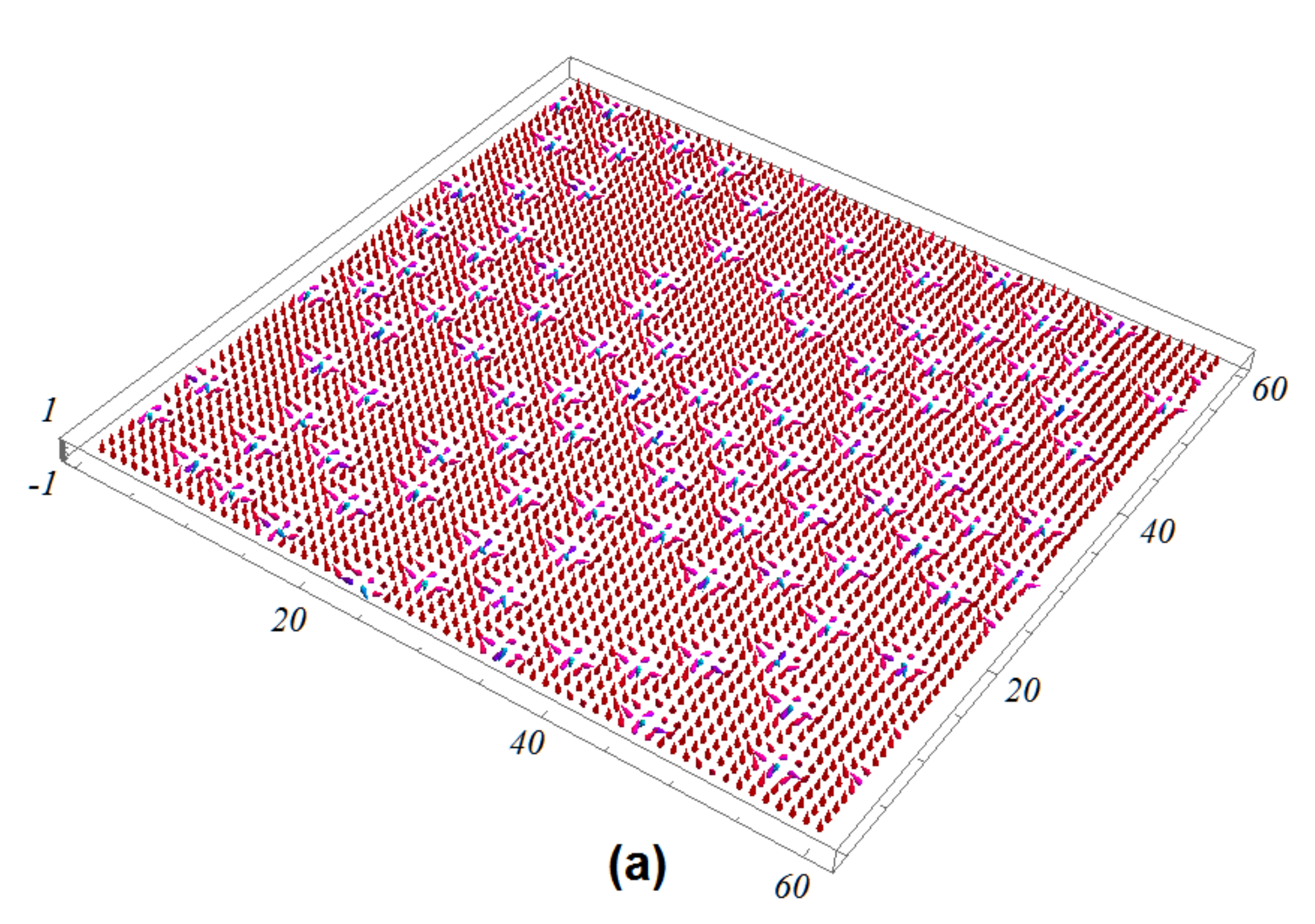}
\includegraphics[scale=0.30]{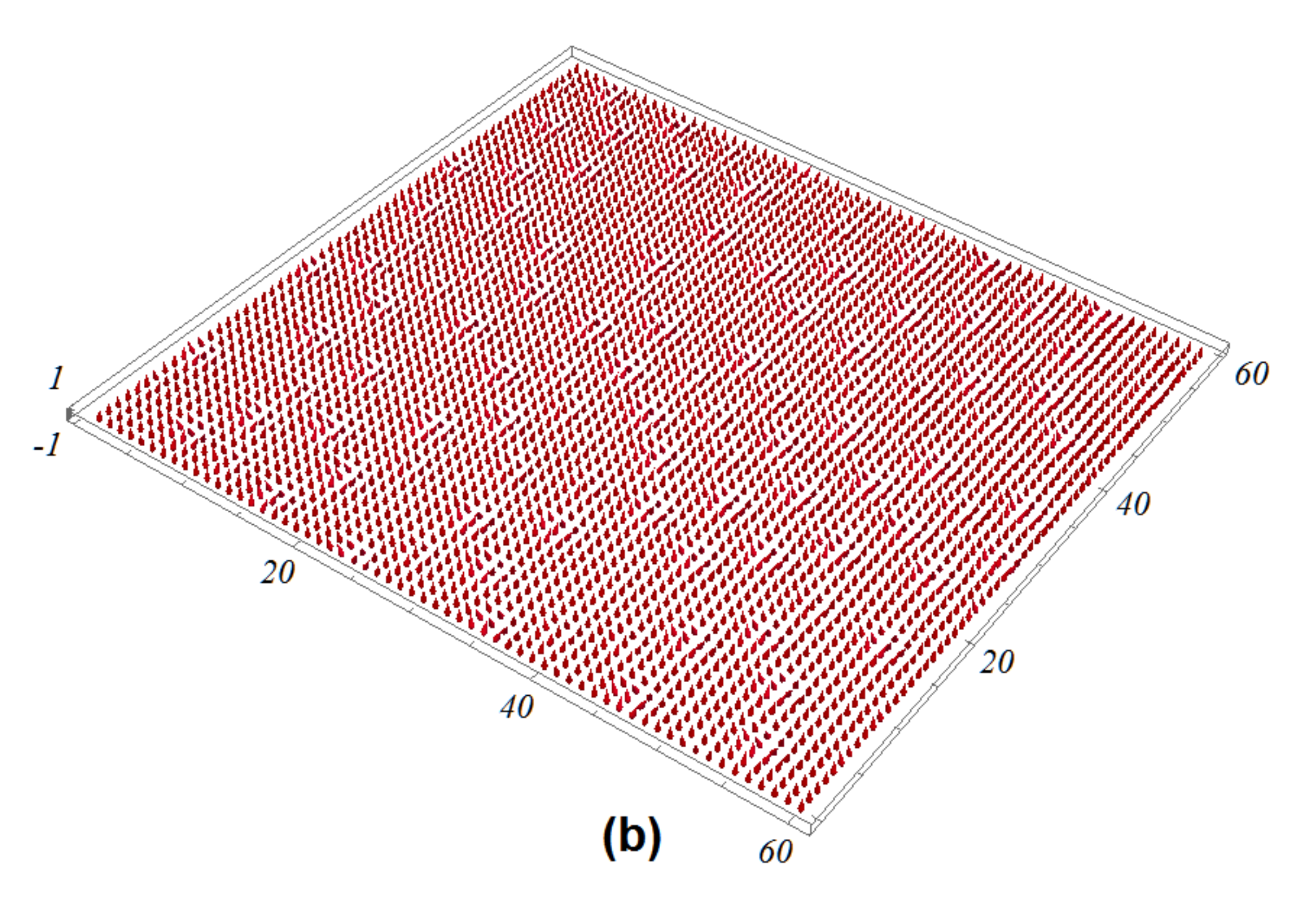}
\end{center}
\vspace{10pt} \caption{ $H=1$: 3D view of the GS configuration of (\textbf{a}) the magnetic interface,  (\textbf{b}) the second magnetic layers. Other parameters: $J^{m}=J^{f}=1$, $J^{2m}=-0.4$, $J^{2f}=-0.4$,
  $J^{mf}=-1.25$.} \label{fig8} \vspace{10pt}
\end{figure}

Let us show now the effect of $H$. In the case of zero frustration, skyrmions
disappear with an external field larger than $H=0.4$\cite{sharafullin2019dzyaloshinskii}. In the case when we take
into account the negative interaction between $NNN$ neighbors, the
skyrmion structure is stable with an external field up to $H=1.0$ (see
Fig \ref{fig8}).  The spins are almost aligned in the direction of the field.

\begin{figure}[h]
\vspace{10pt}
\begin{center}
\includegraphics[scale=0.21]{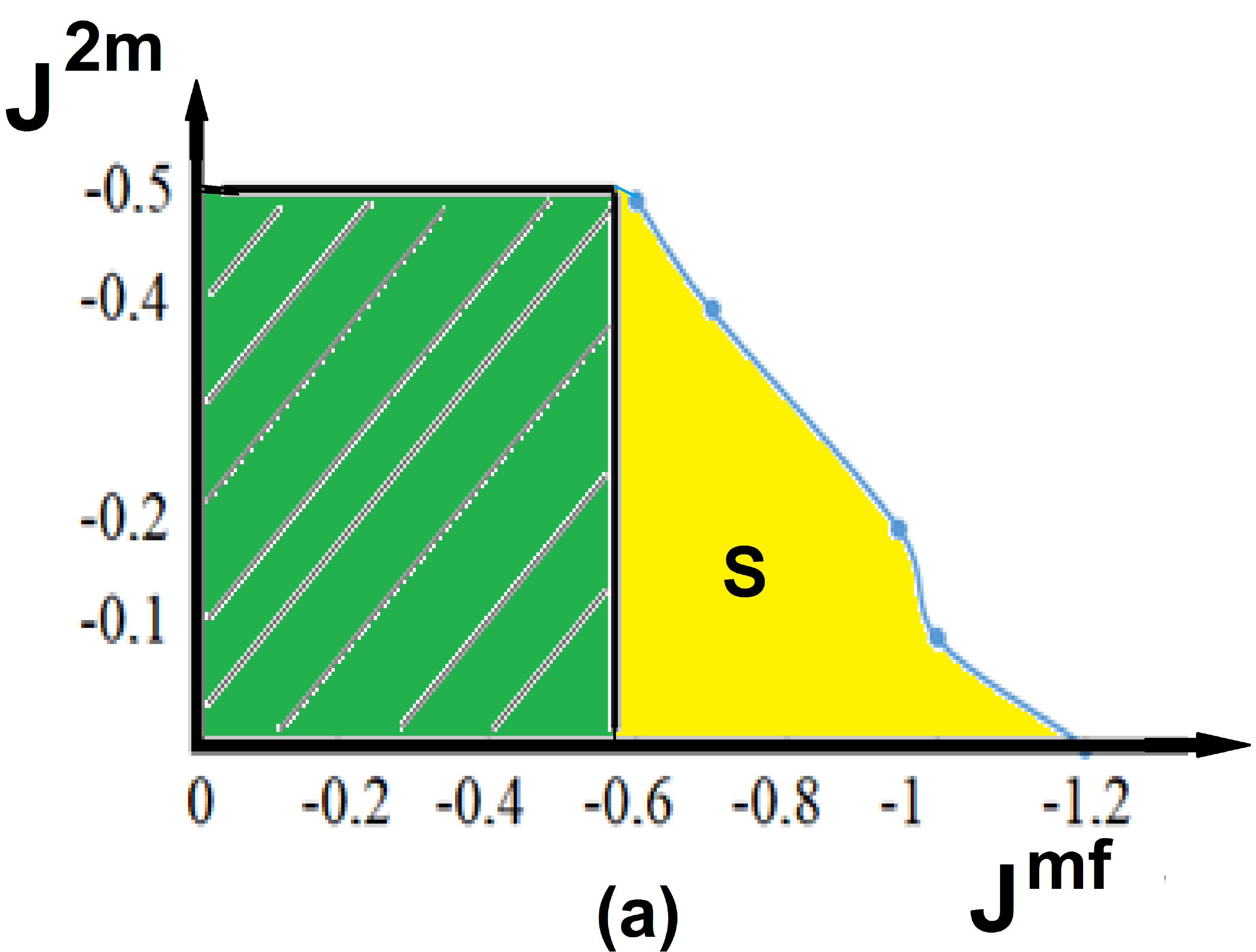}
\includegraphics[scale=0.18]{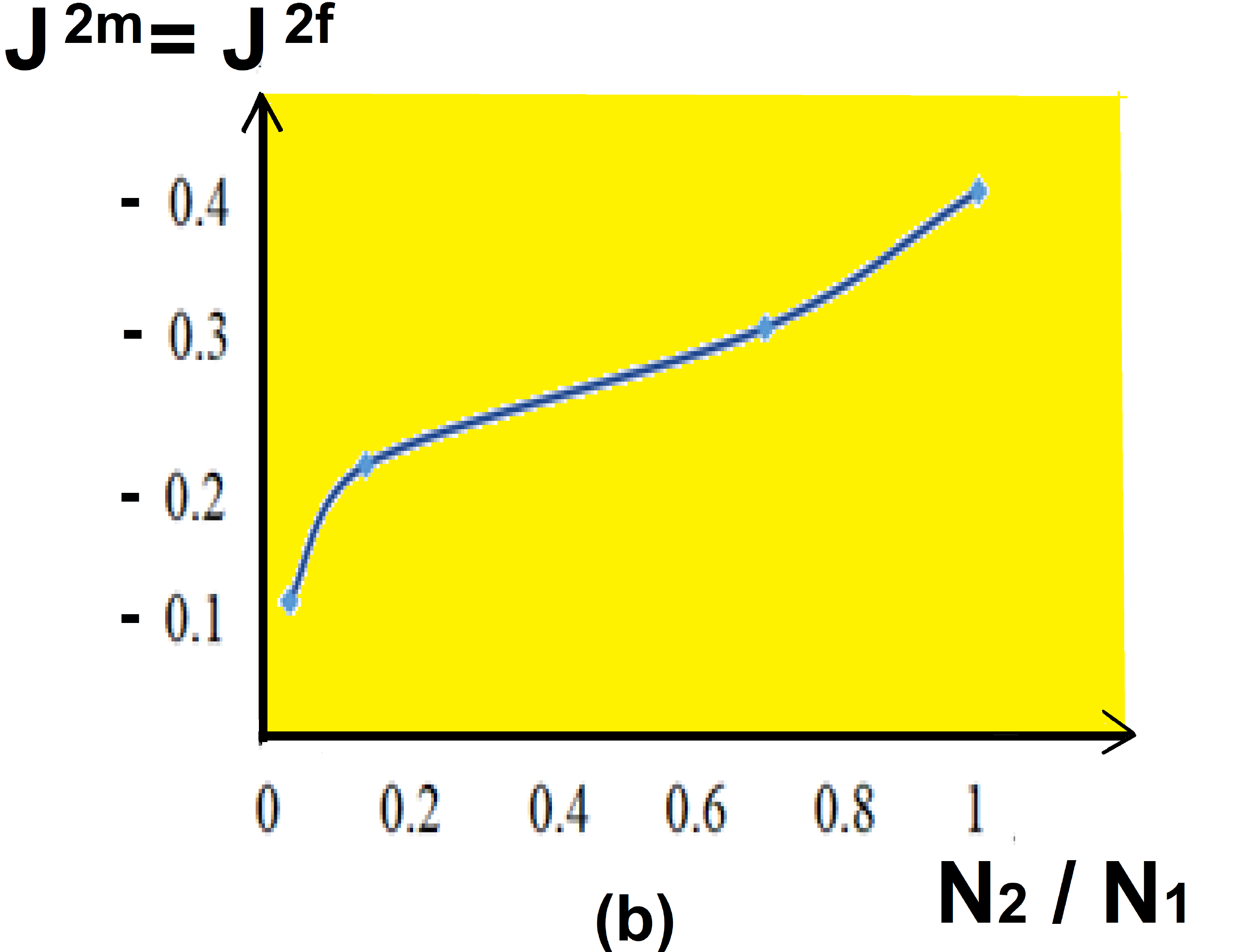}
\end{center}
\vspace{10pt} \caption{(\textbf{a}) Phase diagram in $J^{2m}-J^{mf}$ plane, for the case $J^m=J^f=1$, $J^{2f}=J^{2m}$, $H=0$. The skyrmion phase is indicated by S (the yellow region). See text for comments ; (\textbf{b}) Dependence of the ratio of the number of skyrmions on the interior layer $N_2$  to that on the interface layer $N_1$.} \label{ffig9}
\vspace{10pt}
\end{figure}

The phase diagram in $J^{2m}-J^{mf}$ plane, for the case $J^m=J^f=1$, $J^{2f}=J^{2m}$, $H=0$ is shown in Fig. \ref{ffig9}a.  We can see that in region  $J^{mf}\in [0,-0.6]$
skyrmions are not formed at any value of $J^{2m}$. In region $J^{mf}\in [-0.6,-1.1]$
skyrmions are formed at non-zero values of $J^{2m}$.  The smaller $J^{mf}$
 the larger values of  $J^{2m}$  should be for the formation of skyrmions at
the interface. With $J^{mf}= -1.2$ skyrmions are formed without frustration at
zero values of $J^{2m}$. When we introduce frustration in the magnetic layers at
magneto-ferroelectric interaction  $J^{mf}= -1.2$, skyrmions form a perfect crystalline
structure.  Figure \ref{ffig9}b shows the dependence of the ratio of the number of skyrmions on the interior layer $N_2$  to that on the interface layer $N_1$.  We see that as the frustration becomes stronger the ratio $N_2/N_1$ tends to 1.

Let us discuss about some theoretical observations of skyrmions in frustrated magnets \cite{leonov2017edge, lin2016ginzburg, hayami2016bubble, hayami2016vortices, lin2016magnetic, batista2016frustration, yuan2017skyrmions,sutcliffe2017skyrmion,DiepSD}. Each of these works used a different model, so the comparison is impossible.  However, all shows very similar skyrmion textures. For experiments, a lot of observations have been made in various magnetic materials \cite{zhang2017skyrmion,muhlbauer2009skyrmion,yu2010real,heinze2011spontaneous,romming2013writing,du2015edge,jiang2015blowing, leonov2016chiral,leonov2016properties,woo2016observation,jiang2017direct,litzius2017skyrmion,woo2017spin}, in multiferroic materials \cite{seki2012observation}, in ferroelectric materials \cite{nahas2015discovery}, in semiconductors \cite{kezsmarki2015neel} and in helimagnets \cite{muhlbauer2009skyrmion,yu2010real}.  Here again, each real material corresponds to a particular microscopic mechanism, the comparison is not simple. However, one observes many similar topological textures.

\section{Skyrmion Phase Transition}\label{MC}

The magnetic transition is driven by the competition between $T$, the DM interaction (namely $<P_{k}>$ ), the field $H$ and the
magnetic texture (skyrmions).  The
stronger $<P_{k}>$ and/or $J^{mf}$, the higher the transition temperature $T_c$ of the skyrmion structure.  As mentioned above strong DM interaction helps  stabilize the skyrmion crystal \cite{YangH2018,ManchonA2015} at the superlattice interface. We use a strong $J^{mf}$ as in the previous section.

We use the Metropolis algorithm \cite{Landau09,Brooks11} to simulate the system at  $T\neq 0$. We perform calculations for systems with different sizes $N\times N\times L$ where $N$ varied from 40 to 100 and the thickness $L$ varied from 2 to 16. It should be noted that changing the lateral size of $N$ does not affect the results on skyrmions shown  in the article. But the influence of the total thickness $L$ of the magnetic and ferroelectric layers is very significant: with an increase in the thickness of the magnetic and ferroelectric layers from $L_m= L_f= 4$ to 8,  skyrmions are formed only near the interface, not on layers far inside.  In most calculations, we used $N=60$ and $L_m=L_f=4$.  With this thickness, skyrmions are observed in the two interior layers as seen in the previous section. Usually, we discard $10^6$ Monte Carlo steps (MCS) per spin for equilibrating the system and average physical quantities over the next $10^6$ MCS/spin. Such long MC times are needed since it has been tested for the skyrmion crystal similar to that of the present model \cite{DiepSD}.

For the ferroelectric layers, the order parameter $M_f(n)$ of layer $n$ is given by
\begin{equation}\label{eq-orpar1}
M_f(n)=\frac{1}{N^{2}}\langle{|\sum_{i\in n} P_{i}^z|}\rangle
\end{equation}
where $ \langle{...}\rangle$ indicates the thermal average.

For the magnetic layers where the spin configuration is not collinear, the definition of an order parameter is not easy. One way to do is to heat the system from a selected GS configuration. At a given $T$, we compare the actual spin configuration observed at the time $t$ with its GS. The comparison is made by projecting that configuration on the GS.   The order parameter of layer $n$ can be thus calculated by
\begin{equation}\label{OP}
M_m(n)=\frac{1}{N^2(t_a-t_0)}\sum_{i\in n} |\sum_{t=t_0}^{t_a} \mathbf S_i (T,t)\cdot \mathbf S_i^0(T=0)|
\end{equation}
where $\mathbf S_i (T,t)$ is the i-th spin at the time $t$, at $T$, and $\mathbf S_i^0 (T=0)$ denotes its orientation at $T=0$. We see that $M_m(n)$ is close to 1 at very low $T$ where each spin is close to its orientation in the GS. At high $T$ where every spin strongly fluctuates,  $M_m(n)$ becomes  zero.

Note that $M_m(n)$  is defined in a similar way as the Edwards-Anderson (EA) order parameter in spin glasses \cite{Mezard}. The EA order parameter is calculated by taking the time average of each spin. When it is frozen at low $T$,  its time average is not zero. At high $T$, it strongly fluctuates with time so that its time average is zero. The EA order parameter is just the sum of the squares of each spin average. It expresses thus the degree of freezing but does not express the kind of ordering.


Note that if the system makes a global rotation during the simulation then $\sum_{t=t_0}^{t_a} \mathbf S_i (T,t)\cdot \mathbf S_i^0(T=0)=0$ for a long-time average.
To check this, the most efficient way to do is to calculate the relaxation time to obtain properties at the infinite time, in the same spirit as the finite-size scaling used to calculate properties at the infinite system size. We have calculated the relaxation time of the 2D skyrmion crystal \cite{DiepSD} using the order parameter defined by Eq. (\ref{OP}). We have found that skyrmions need more than $10^6$ MCS/spin to relax to equilibrium and the order parameter follows a stretched exponential law as in SG for $T<T_c$.

Another way to check the stability of the skyrmion crystal is to count numerically the topological charges around each skyrmion\cite{Koibuchi-Diep2019}. If there is a phase transition, the charge number, which plays the role of an order parameter,  evolves  with $T$ and goes to zero at the phase transition.

The total order parameters $M_m$ and $M_f$ are the sum of the layer order parameters, namely $M_m=\sum_n M_m(n)$ and $M_f=\sum_n M_f(n)$.

\begin{figure}[h]
\vspace{10pt}
\begin{center}
\includegraphics[scale=0.05]{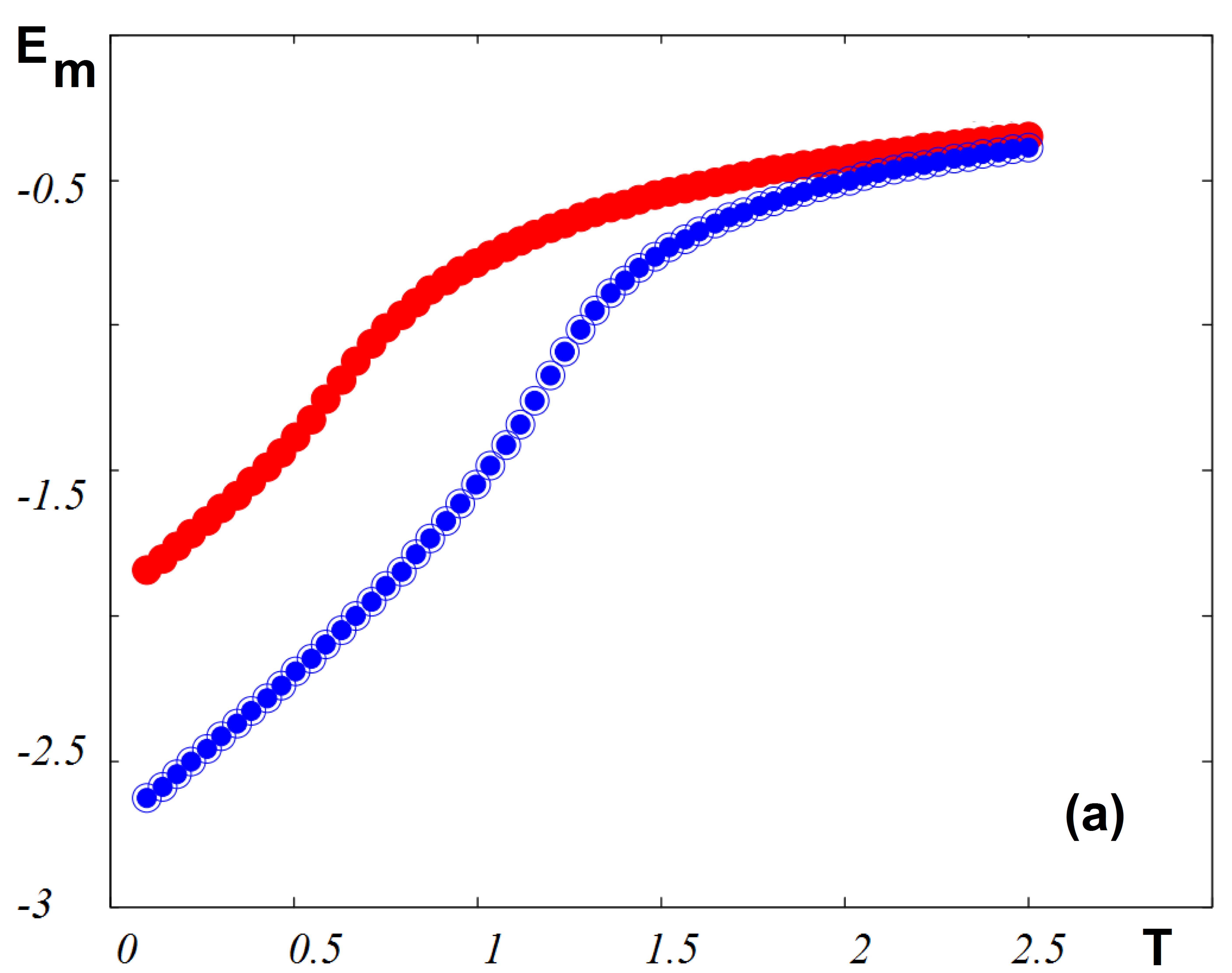}
\includegraphics[scale=0.05]{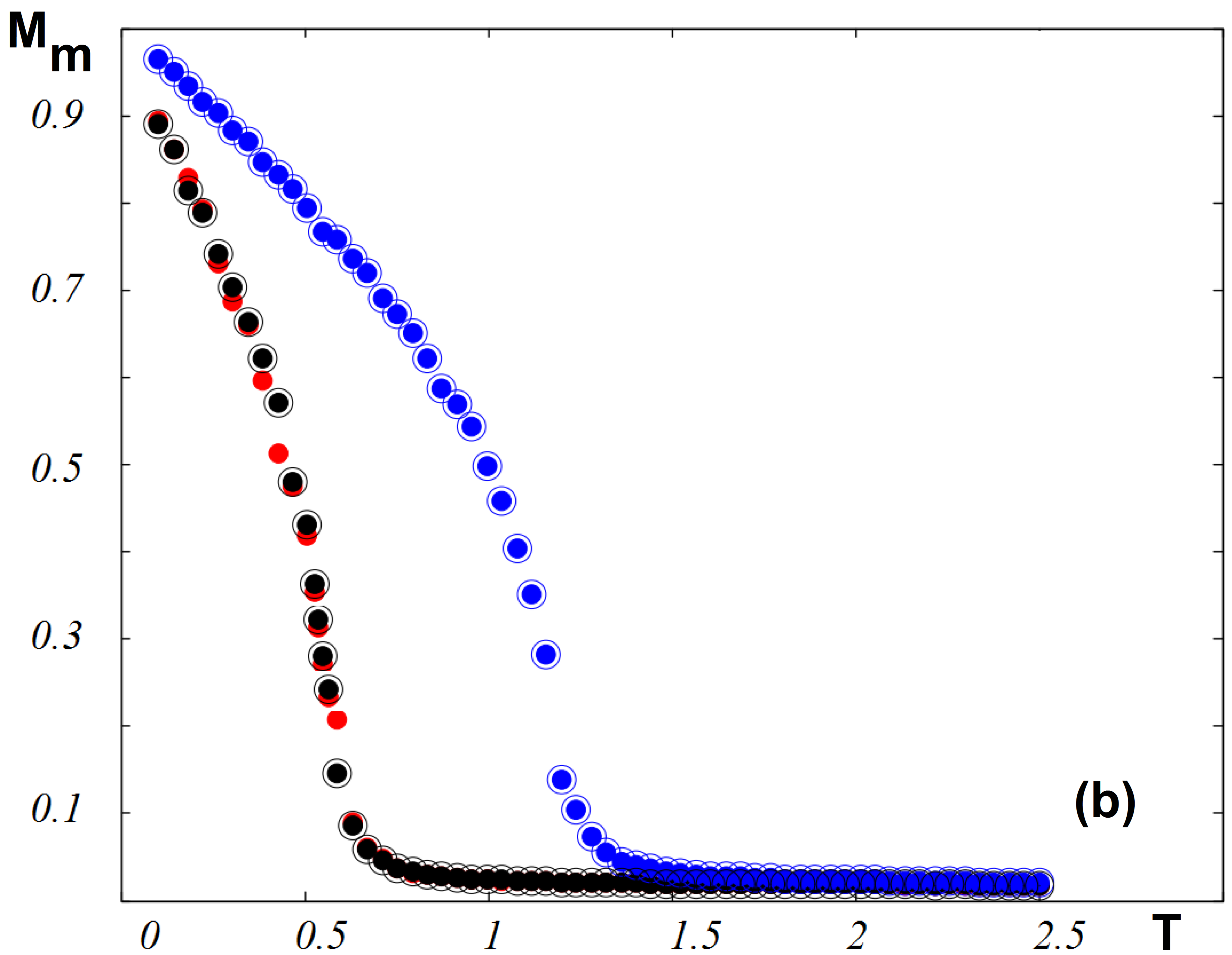}
\end{center}
\vspace{10pt} \caption{(\textbf{a}) Energy of the magnetic films  versus temperature $T$ for $(J^{2m}=J^{2f}=-0.4)$ (red), coinciding with the curve for $(J^{2m}=-0.4, J^{2f}=0)$ (black, hidden behind the red curve). Blue curve is for $(J^{2m}=0, J^{2f}=-0.4)$; (\textbf{b}) Order parameter of the magnetic films versus temperature $T$ for
$(J^{2m}=J^{2f}=-0.4)$ (red), $(J^{2m}=-0.4, J^{2f}=0)$ (black), $(J^{2m}=0, J^{2f}=-0.4)$ (blue). Other used parameters: $J^{mf}=-1.25$, $H=0.25$.} \label{fig9}
\vspace{10pt}
\end{figure}

We display now in Fig.\ref{fig9} the magnetic
energy and the magnetic order parameter vs $T$ in an external magnetic
field, for various sets of NNN interaction.
Note that the phase transition
occurs at the energy curvature changes, namely at the
maximum of the specific heat.
The red curve in Fig.\ref{fig9}a is for both sets $(J^{2m}=J^{2f}=-0.4)$, $(J^{2m}=-0.4, J^{2f}=0)$. The change of curvature takes place at
$T_c^m\simeq
0.60$. It means that the ferroelectric frustration does not affect the magnetic skyrmion transition at such a strong magnetic frustration $(J^{2m}=-0.4)$.
For $(J^{2m}=0, J^{2f}=-0.4)$, namely no magnetic frustration,  the transition takes place at a much higher temperature $T_c^m\simeq
1.25$.

At this stage, we note that the above results are shown in the dimensionless unit: energy in unit of $J^m$ and temperature in unit of $J^m/k_B$.  Our results can be used for materials with different $J^m$. For example, if experimentally one observes that the skyrmion phase transition occurs at $T_c^{exp}=200$ K, we can calculate the effective exchange $J_{eff}$ using for example the mean-field equation
\begin{eqnarray}
T_c^{exp}&=&\frac{2}{3k_B}ZS(S+1)J_{eff}\label{exp1}\\
J_{eff}&=& 34.51\times  10^{-23}   \ \ {\mbox {Joules}}\\
&=& 47.63  \ \ {\mbox {K}}
\end{eqnarray}
where $Z=6$ (simple cubic lattice), $S=1$ (spin magnitude) and $k_B=1.3807 \times 10^{-23}$ Joules/Kelvin have been used.   $J_{eff}$ is a combination of $J^m$,  $J^{2m}$ and $J^{mf}$. Knowing the GS, we can deduce these interactions.  We can calculate then the energy in unit of Joule by multiplying the value of $E_m$ in Fig. \ref{fig9}a by the value of $J^m$.  Unfortunately, there is at the time being no experimental energy measurement for comparison.

The magnetic order parameters shown in Fig. \ref{fig9}b confirm
the skyrmion transition temperatures seen by the curvature change of the energy in Fig.\ref{fig9}a.

\begin{figure}[h]
\vspace{10pt}
\begin{center}
\includegraphics[scale=0.05]{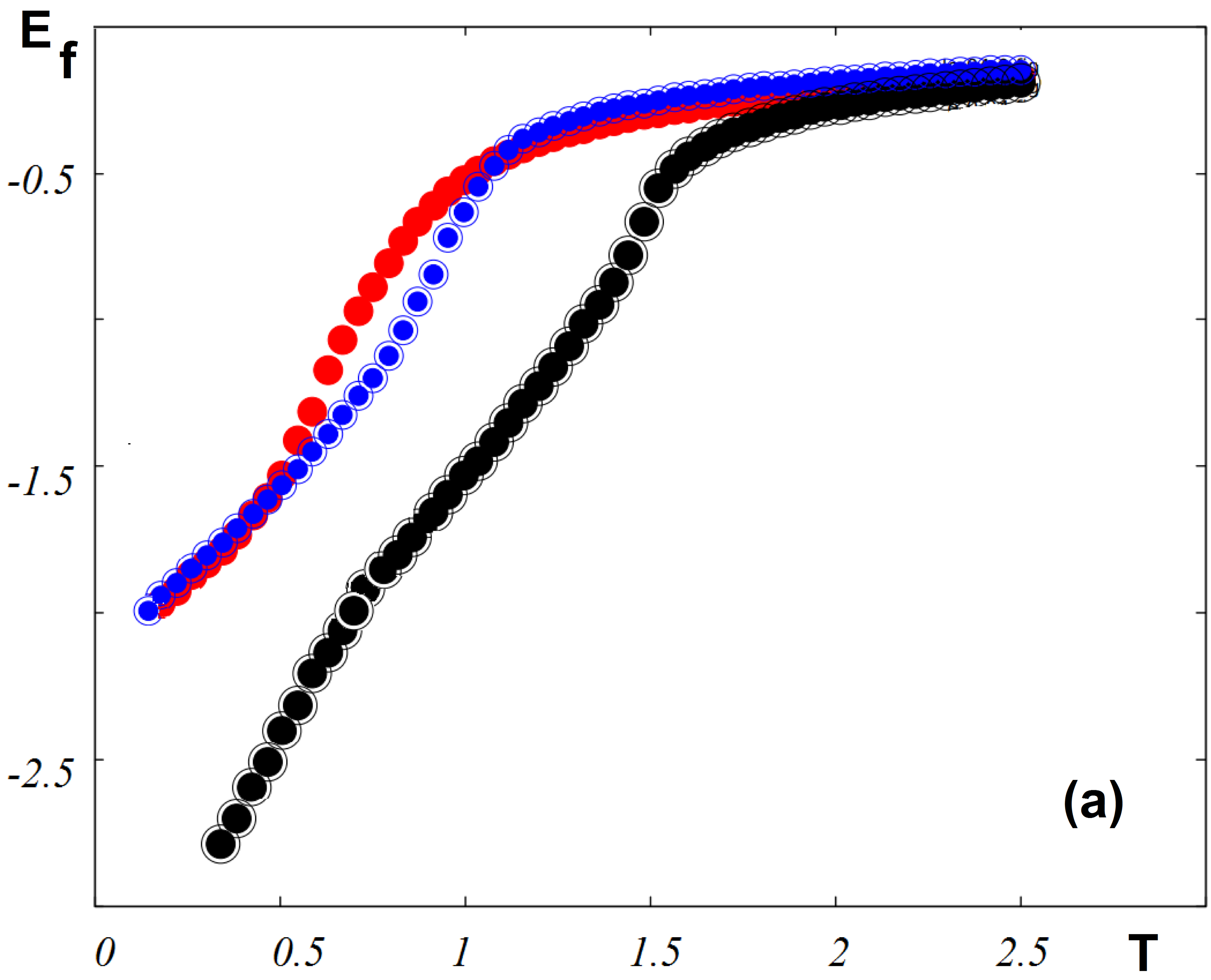}
\includegraphics[scale=0.05]{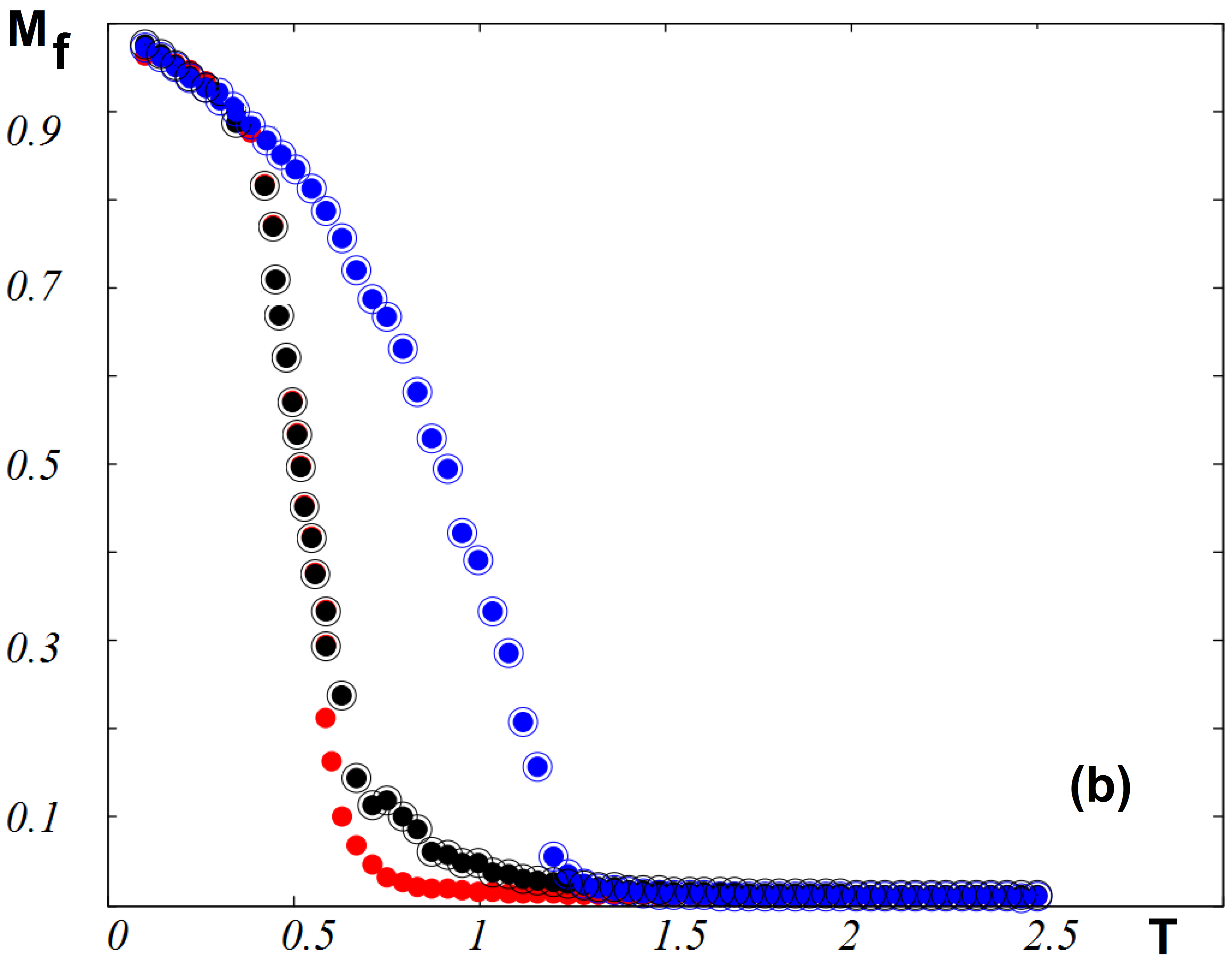}
\end{center}
\vspace{10pt} \caption{(\textbf{a}) Energy of the ferroelectric films versus temperature $T$ for $(J^{2m}=J^{2f}=-0.4)$ (red), $(J^{2m}=-0.4, J^{2f}=0)$ (black), ($J^{2m}=0, J^{2f}=-0.4$) (blue),
(\textbf{b}) Order parameter of the ferroelectric films versus temperature $T$ for
$(J^{2m}=J^{2f}=-0.4)$ (red), $(J^{2m}=-0.4, J^{2f}=0)$ (black), $(J^{2m}=0, J^{2f}=-0.4)$ (blue). Other used parameters: $J^{mf}=-1.25$, $H=0.25$.} \label{fig10} \vspace{10pt}
\end{figure}

We show in Fig. \ref{fig10}a  the ferroelectric energy and ferroelectric order parameters for the same sets of frustration parameters:
($J^{2m},J^{2f}=-0.4$),  ($J^{2m}=-0.4$, $J^{2f}=0$), ($J^{2m}=0$, $J^{2f}=-0.4$). As seen, the first and second sets where the magnetic frustration is strong give respectively $T_m^f\simeq 0.60$, 0.90. It means that the ferroelectric frustration which does not affect the skyrmion transition, strongly affects the ferroelectric transition.
While, the third set with no magnetic frustration ($J^{2m}=0$) gives the transition at $T_c^m\simeq 1.50$.
Figure \ref{fig10}b shows the ferroelectric order parameters for the three sets of NNN interactions shown in Fig. \ref{fig10}a. These curves confirm the transition temperatures given above.

\section{Conclusion}\label{Concl}

We have studied in this paper the effect of the NNN interactions in both magnetic and ferroelectric layers of a magneto-ferroelectric superlattice. A Dzyaloshinskii-Moriya (DM) interaction was assumed for the  magneto-ferroelectric interface coupling.

We found  the formation of a skyrmion crystal in the GS under an applied magnetic field in a large region of parameters in the space ($J^{2m},J^{mf})$.
As expected, the magnetic frustration enhances the creation of skyrmions while the ferroelectric frustration when strong enough destabilizes skyrmions if the magnetic frustration is weak.

We have studied the phase transition of the skyrmion crystal by the use of Monte Carlo method. Skyrmions have been shown to be stable at finite temperatures. While the magnetic frustration helps enhance the creation of skyrmions, it reduces the transition temperature considerably.

The existence of very stable skyrmions confined at the magneto-ferroelectric
interface at  finite $T$ is very interesting and may have potential applications in spintronics.   Many applications using skyrmions have been mentioned in the
Introduction. As a last remark, let us mention that the present magneto-ferroelectric superlattice model can be used in the case of magnetic monolayer or bilayer to study the dynamics of the skyrmions driven by a spin-polarized current or by a spin-transfer torque. Due to the small thickness, the skyrmions created by the interface are well confined as in 2D.   Our model is therefore suitable for creating skyrmion pinning by using an electric field acting on the ferroelectric polarizations. This subject is our future investigation. 





\bibliographystyle{elsarticle-num}
\bibliography{bibliography}

\end{document}